%% file: pcocal.tex
\documentclass[11pt,twoside,letterpaper]{article}
\usepackage{latexsym}
\usepackage{amsmath}
\usepackage{dcolumn}
\usepackage{enumitem}
\usepackage{authblk}

\usepackage[margin=1in]{geometry}

\usepackage{graphicx}
\usepackage{amssymb}
\usepackage{bm}
\usepackage{color}
\usepackage{soul}
\usepackage{hyperref}
\usepackage{enumerate}
\usepackage{stackengine}
\usepackage{algorithm}
\usepackage[noend]{algpseudocode}
\usepackage[figuresright]{rotating}
\usepackage{orcidlink}

\definecolor{green1}{rgb}{0.0, 0.5, 0.0}

\input defs.tex

\begin{document}

\title{The Parallel Compact Object CALculator: An Efficient General Relativistic Initial Data Solver for Compact Objects} 

\author[1]{Lambros Boukas}
\author[2,3,4]{Antonios Tsokaros}
\author[5]{K{\= o}ji Ury{\= u}}

\affil[1]{LAB Scientific Computing Inc., Miami, FL 33176, USA}
\affil[2]{Department of Physics, University of Illinois at Urbana-Champaign, Urbana, IL 61801, USA}
\affil[3]{National Center for Supercomputing Applications, University of Illinois at Urbana-Champaign, Urbana, IL 61801, USA}
\affil[4]{Research Center for Astronomy and Applied Mathematics, Academy of Athens, Athens 11527, Greece}
\affil[5]{University of the Ryukyus, Department of Physics, Senbaru, Nishihara, Okinawa 903-0213, Japan}


\maketitle

\begin{abstract}
Every numerical general relativistic investigation starts from the solution of the initial value
equations at a given time. Astrophysically relevant initial values for different systems lead to 
distinct set of equations that obey specific assumptions tied to the particular problem. Therefore
a robust and efficient solver for a variety of strongly gravitating sources is needed. In this
work we present the OpenMP version of the Compact Object CALculator (COCAL) 
on shared memory processors. We performed extensive profiling of the
core COCAL modules in order to identify bottlenecks in efficiency which we addressed. Using modest
resources, the new parallel code achieves speedups approximately one order of magnitude relative to 
the original serial COCAL code, which is crucial for 
parameter studies of computationally expensive systems, such as magnetized neutron stars, as well as its
further development towards more realistic scenarios.
As a novel example of our new code we compute a binary quark system where each companion
has a dimensionless spin of $0.43$ aligned with the orbital angular momentum.

\end{abstract}

\section{Introduction}
\label{sec:intro}

Gravitational wave astronomy was launched in 2015 with the first-ever
gravitational wave detection of the inspiral and merger from a binary black hole
system, as reported by the LIGO scientific collaboration --- event
GW150914 \cite{Abbot2016-GW-detection-prl}. Two years later the simultaneous
detection of gravitational waves from an inspiraling binary neutron star
system, event GW170817, and its postmerger emission of electromagnetic
radiation spurred the era of multimessenger astronomy
\cite{GW170817prl,Kolzova2017circ,Abbott2017b,Abbott2017d_etal,Abbott2017c}.
Merging binary neutron stars and black hole--neutron stars  are not only
important sources of gravitational radiation, but also promising candidates for
coincident electromagnetic counterparts, which could give new insight into
their sources. 
To  understand these observations and, in particular, to understand the physics
of matter under extreme conditions, it is crucial to compare them to
predictions from theoretical modeling, which, due to the complexity of the
underlying physical phenomena, are largely numerical in nature. 

Numerical modeling of strongly gravitating systems 
typically involves two steps: First, the ab initio calculation
of the initial values that describe the astrophysical system at a given time,
and second, the evolution of these initial data in order to describe it at
later times. These two general relativistic calculations involve two distinct
codes. One that solves the elliptic problem to find out the initial data and
another that solves the hyperbolic one that simulates their evolution.  
For that reason we have developed the
Compact Object CALculator (COCAL) code which is a {\it serial} code that solves for any compact object in
equilibrium or quasiequilibrim in a variery of settings and/or mathematical
formulations.  In particular COCAL has been used to calculate: (i) Rotating
neutron stars and quark stars (axisymmetric or triaxial, uniformly or
differentially rotating)
\cite{Huang08,Uryu2016a,Uryu2016b,Uryu2017,Zhou2017xhf,Zhou2019hyy}.  (ii)
Binary black holes in quasicircular orbits
\cite{Uryu2012,Tsokaros2012,Uryu:2012b}.  (iii) Binary neutron stars and quark
stars (irrotational or spinning) in quasicircular orbits
\cite{Tsokaros2015,Tsokaros2018}.  (iv) Magnetized rotating neutron stars with
generic mixed poloidal and toroidal magnetic fields and a magnetosphere 
\cite{Uryu2014,Uryu2019,Uryu:2023lgp}.
(v) Self-gravitating, tilted  black hole-disks (where the angular momentum of the disk is tilted
with respect to the angular momentum of the black hole) \cite{Tsokaros2018a}.  
The main characteristics of the COCAL code is the use of
finite differences and a Green's function approach as first developed in
\cite{Uryu00} for neutron stars and in \cite{Tsokaros2007} for black holes to
achieve a convergent solution through a Picard type of iteration, known as the
Komatsu-Eriguchi-Hachisu (KEH) method \cite{Komatsu89,Komatsu89b}.  The field
equations are solved in spherical coordinates in multiple patches and a smooth
solution is obtained everywhere through boundary surface integrals. For the
fluid equations surface fitted coordinates are being implemented that allow
accurate representation of the neutron star surface which is important in order
to impose boundary conditions.

Similar to COCAL, there exist a number of other codes whose purpose is
to solve for initial data in general relativity. For binary as well as for
single compact objects codes these include the TwoPunctures
\cite{Ansorg:2004ds}, LORENE \cite{lorene}, KADATH/FUKA
\cite{kadath,Grandclement09,Papenfort2021}, SGRID \cite{Tichy:2009}, Spells
\cite{Pfeiffer:2002wt}, Elliptica \cite{Rashti2021}, and NRPyElliptic
\cite{Assumpcao2021}.  Each code uses each own numerical methods, and depending
on the problem, it may use different mathematical formulations too.

Compact systems (i)--(v) employ different mathematical formulations that
nevertheless share common characteristics. For example, one well known method
for the calculation of initial data \cite{Tsokaros2022r} is the
Isenberg-Wilson-Mathews (IWM)  \cite{Isenberg08,Wilson89,Wilson95,Wilson96}
formulation where the spatial three-dimensional metric is assumed to be
conformally flat and the trace of the extrinsic curvature, zero. 
This method is applied from rotating neutron/quark stars
\cite{Huang08,Zhou2017xhf} to binary black holes
\cite{Uryu2012,Tsokaros2012,Uryu:2012b} and binary neutron stars
\cite{Tsokaros2015,Tsokaros2018} and results to five elliptic equations, i.e.,
a truncated system of the Einstein field equations. Notwithstanding the
approximate nature of the method, its robustness and accuracy (especially for
neutron stars) make the IWM method a workhorse in initial data
construction. On the other hand, more  intricate methods have been developed
that can treat the {\it full} Eistein system both for neutron stars and black
holes. In \cite{Uryu2016a}  rotating axisymmetric or triaxial neutron stars
have been computed using the waveless method \cite{Shibata04a,Bonazzola:2003dm}, while in
\cite{Uryu2014,Uryu2019,Uryu:2023lgp} this method is employed to calculate generic
magnetized equilibria with mixed poloidal and toroidal magnetic fields as well as
with a force-free magnetosphere. In
\cite{Uryu2006,Uryu:2009ye} the same method has been used to compute binary
neutron stars in quasicircular orbits. This method is referred as the ``waveless''
formulation. For spacetimes that contain a black hole
a new method has been developed in \cite{Tsokaros2018a} that solves the
complete initial data equations. Contrary to the IWM method, the
formulations that try to address the whole Einstein system \cite{Shibata04a,Tsokaros2018a}
result to a significantly larger number of elliptic equations. For neutron
star spacetimes \cite{Shibata04a} we have fourteen nonlinear Poisson-like equations, with 
four of them resulting from imposing gauge conditions. For black hole spacetimes 
we have a total of seventeen equations with four of them resulting from imposing 
gauge conditions and another three for an additional decomposition of the
extrinsic curvature \cite{Tsokaros2018a}. The inclusion of magnetohydrodynamics
will increase the total number of elliptic equations to be solved even more. Therefore it is
evident that in order to be able to study these astrophysically realistic systems in a 
systematic manner, an efficient parallelization scheme for the COCAL code needs to be developed.

In this paper we present the OpenMP version of the COCAL code which we call
the Parallel Compact Object CALculator or PCOCAL for short.
For the first time we have analyzed systematically the whole serial COCAL code and identified the bottlenecks both
for single objects (e.g. rotating neutron stars) as well as for binaries (e.g. binary neutron stars)
using either the IWM or the waveless formulation. 
In order to implement the OpenMP directives for parallelization 
we had to revise and sometimes completely rewrite many of the original subroutines.
Using modest resources of 30-40 cores both the single rotating
neutron star module as well as the binary neutron star one achieve a speedup of an order of magnitude. In the
binary case we observed a smaller speedup than in the single star case due mainly to the communication between 
different coordinate systems which is absent in the latter calculation.
As an application of our new code we computed for the first time a spinning binary quark star system
where each companion has a dimensionless spin of $0.43$ aligned with the orbital angular momentum.

In the equations presented, Greek indices are taken to run from 0 to 3 while Latin indices from 1 to 3.
We use a signature $(-,+,+,+)$ for the spacetime line element, and a system of units
in which $c=G=M_\odot =1$ (unless explicitly shown).

\section{The initial value equations}
\label{sec:eqs}

In this section, we summarize the 3+1 decomposition of Einstein's equations 
for two representative problems that will be discussed below: (i) single rotating neutron stars in the
waveless formalism \cite{Shibata04a,Uryu2016a} and (ii) spinning binary neutron stars in the
IWM formalism \cite{Tsokaros2015,Tsokaros2018}. While details on these methods can be found in
the aforementioned papers, here we will give a broad overview of the equations that need to be
solved in order to discuss the time footprint of each one in the iteration procedure, the
possible bottlenecks with the parallelization as well as the optimizations we performed.

We assume that a spacetime $\cal M$ is foliated 
by a family of spacelike hypersurfaces $\Sigma_t$, 
parametrized by a time coordinate $t\in {\mathbb R}$ as 
${\cal M} = {\mathbb R} \times \Sigma_t$.  
The future-pointing unit normal one form to $\Sigma_t$, $n_\alpha = -\alpha\na_\alpha t$, 
is related to the generator of time translations $t^\alpha$ as 
\beq
t^\alpha := \alpha n^\alpha + \beta^\alpha, 
\eeq
where $t^\alpha \na_\alpha t = 1$. Here, 
$\alpha$ and $\beta^\alpha$ are, respectively, the lapse function and the shift vector
 which is spatial, $\beta^\alpha \na_\alpha t=0$.  
The projection tensor $\gamma^{\alpha}_{\ \,\beta}$  to $\Sigma_t$  is introduced as 
$\gamma^{\alpha}_{\ \,\beta} := g^\GA_{\ \,\GB} + n^\alpha n_\beta$.  
The induced spatial metric $g_{ij}$ on $\Sigma_t$ is the 
projection tensor restricted to it.  
Introducing a conformal spatial geometry with 
spatial metric $\TDD{\GG}{i}{j}=\GC^{-4}\GG_{ij}$, 
where $\psi$ being the conformal factor,
we can write the line element on a chart $\{t,x^i\}$ of $\Sigma_t$ as
\beq
ds^{2}=-\alpha^{2}dt^{2}+\psi^{4}\TDD{\GG}{i}{j} (dx^i+\beta^i dt)(dx^j+\beta^j dt).
\eeq
The conformal rescaling is determined from a condition 
$\tilde{\GG} =f$, where $\tilde{\GG}$ and $f$ are determinants 
of the rescaled spatial metric $\TDD{\GG}{i}{j}$ and the flat metric 
$f_{ij}$, respectively.  
We denote by $h_{ij}$ and $h^{ij}$ the differences 
between the conformal and the flat metric 
\beq
\TDD{\GG}{i}{j} := f_{ij}+h_{ij}\,,\qquad  \TUU{\GG}{i}{j} := f^{ij}+h^{ij}. 
\eeq

The extrinsic curvature of each slice $\Sigma_t$ is defined by 
\beqn
K_{ij} = -\frac{1}{2}\mathcal{L}_{\Bn}\GG_{ij}
\label{eq:Kab}
\eeqn
where $\mathcal{L}_\Bn$ is the Lie derivative along the normal vector on $\Sigma_t$.  
Hereafter, we denote the trace of $K_{ij}$ by $K$, and 
the tracefree part of $K_{ij}$ by $A_{ij} :=  K_{ij} -\frac{1}{3} \GG_{ij} K $. 
We define the conformal extrinsic curvature $\TDD{A}{i}{j}=\GC^{-4} A_{ij}$, 
similar to the conformal spatial metric.

In this paper, we consider perfect-fluid spacetimes whose stress-energy tensor is written as 
\beq
T^{\GA\GB} = (\epsilon+p)u^\alpha u^\beta + p g^{\GA\GB}, 
\label{eq:Tab}
\eeq
where $\epsilon$ is the energy density and $p$ the pressure as measured by the comoving observer, i.e.
an observer with 4-velocity $u^\alpha$.   As discussed in \cite{Uryu2014,Uryu2019}, in the case where 
we have magnetized neutron stars the total stress-energy tensor in addition to the perfect-fluid  contribution, Eq. (\ref{eq:Tab}), 
will include the contribution of the magnetic field. We decompose the Einstein's equations 
$\mathcal{E}_{\GA\GB} := G_{\GA\GB} -8\pi T_{\GA\GB}=0$ along the hypersurface as well as along its 
normal $n^\alpha$ as follows, 
\beqn
& \mathcal{E}_{\GA\GB} n^\alpha n^\beta \,=\,0,&  \label{eq:HamC} \\
& \mathcal{E}_{\GA\GB} \GG^\GA_{\ \,i} n^\beta \,=\,0,&  \label{eq:MomC} \\
& \mathcal{E}_{\GA\GB} \left(\gamma^{\GA\GB}+\frac{1}{2} n^\alpha n^\beta\right) \,=\,0,&  \label{eq:trG}   \\
& \mathcal{E}_{\GA\GB} \left(\GG^\GA_{\ \,i} \GG^\GB_{\ \,j} -\frac{1}{3} \GG_{ij} \gamma^{\GA\GB}\right)\,=\,0,& \label{eq:trfreeG}  
\eeqn
which correspond, respectively, to the Hamiltonian constraint, the momentum constraint, 
spatial trace part (combined with the Hamiltonian constraint), 
and spatial tracefree part.  

As shown in \cite{Shibata04a}, the above set of field equations 
(\ref{eq:HamC})-(\ref{eq:trfreeG}) are reduced to elliptic (Poisson) 
equations for $\{ \psi, \TD{\GB}{i}, \alpha\psi, h_{ij} \}$ respectively, 
as follows, 
\beqn
\flap \psi &=& \mathcal{S}_{\rm H}, \label{eq:psi} \\
\flap \TD{\GB}{i} &=& \mathcal{S}_i, \label{eq:shift}  \\
\flap (\alpha\psi) &=& \mathcal{S}_{\rm tr}, \label{eq:lapse} \\
\flap h_{ij} &=& {\cal S}_{ij}, \label{eq:hab}
\eeqn
where $\flap$ is the flat metric (in arbitrary coordinates) Laplacian, defined by $\flap = f^{ij}\fD_i\fD_j$. 
The sources $\mathcal{S}_{\rm H}$, $\mathcal{S}_i$, $\mathcal{S}_{\rm tr}$, and ${\cal S}_{ij}$ \cite{Tsokaros2018a}
depend {\it nonlinearly} on the unknown potentials $\{ \psi, \TD{\GB}{i}, \alpha\psi, h_{ij} \}$ 
and are written in \ref{sec:st}. 
In the COCAL code we use the Cartesian components of the elliptic equations (\ref{eq:psi})-(\ref{eq:hab}),
on spherical grids (see Section \ref{sec:nm}).
Equations (\ref{eq:psi})-(\ref{eq:hab}) must be supplied with conditions on the
boundary of our computational region. Since we will consider only isolated single or binary
stars, the boundary conditions for the gravitational equations will be only at
spatial infinity, where we impose asymptotic flatness, i.e.
\beq
\lim_{r\rightarrow\infty} \GA = \lim_{r\rightarrow\infty} \GC = 1\,,\qquad 
\lim_{r\rightarrow\infty} \GB^i = \lim_{r\rightarrow\infty} h_{ij} =  0  .
\label{eq:bcpsal}
\eeq

\subsection{Rotating neutron star in the waveless formalism}
\label{sec:rns}

Coordinate gauge conditions such as 
the maximal slicing and the generalized Dirac gauge 
\beqn
& K=0 &   \label{eq:maximal_slicing}  \\
& H^i = 0 &   \label{eq:Dirac_gauge}
\eeqn
are assumed for rotating neutron stars. 
As described in \ref{sec:st}, $H^i = \fD_j \tGG^{ij}$.
These conditions simplify
Eqs. (\ref{eq:HC})-(\ref{eq:tfdKijdt}) significantly. In particular
$\tRs_{ij}^{\li}=0$. 
The asymptotic behavior of the metric potentials becomes 
a Coulomb type fall off,
\beqn
\psi-1 = O(r^{-1}),\qquad &  \alpha-1 = O(r^{-1}),\qquad    \\
\beta^i = O(r^{-2}),\qquad & h_{ij}  = O(r^{-1}).\qquad    
\eeqn
Our choice of $\beta^a$ is the shift in an (asymptotically) inertial frame.  

The Bianchi identity implies $\nabla_\GA T^{\GA\GB}=0$. Under the assumption
of rest-mass conservation $\nabla_\GA(\GR u^\GA)=0 $, and an isentropic law $\nabla_\GA s =0$,
it leads to the relativistic Euler equation
\beq
u^\GB \GO_{\GB\GA} = 0 \ ,  \qquad\mbox{where}\qquad  \GO_{\GA\GB} := \nabla_\GA(hu_\GB) - \nabla_\GB(hu_\GA)
\label{eq:reeq}
\eeq
is the relativistic vorticity and $h:=(\GE+p)/\GR$ the relativistic specific enthalpy.
We consider two cases of single rotating stars, both of which are described by a 4-velocity 
\beq
u^\GA = u^t k^\GA \ .
\label{eq:4velrns}
\eeq
Here $k^\GA = t^\GA + \Omega \GP^\GA$ is the helical symmetry vector, $\Omega$ the constant
angular velocity of the star and $\GP^\GA$ is the generator of rotational symmetry.
For uniformly rotating stars it is $u^\GB \GO_{\GB\GA} = - u^t \nabla_\GA (h u_\GB k^\GB)$.

\subsubsection{Axisymmetric rotating neutron stars}
\label{sec:rnsaxi}

For stationary and axisymmetric systems, we  impose  time symmetry 
on both the three dimensional metric as well as the extrinsic curvature
\beq
\mathcal{L}_{\Bt} \GG_{ij} \,=\, \mathcal{L}_{\Bt} K_{ij} \,=\, 0  .
\label{eq:axitime}
\eeq 
Note that we do not explicitly impose the axisymmetry on our formulation.  
Under these asumptions 
$\tilde{u}_{ij}=0$ in Eq. (\ref{eq:MC}) and $\mathcal{L}_{\GA\Bn} (\cdots) = - \mathcal{L}_{\Bbe} (\cdots)$
for Eqs. (\ref{eq:tfdKijdt}). Note that the last terms involve second order derivatives of the shift vector $\TU{\GB}{i}$.

The stationarity condition for the fluid variables 
\beq
\mathcal{L}_{\Bt} (h u_\GA) \,=\, \mathcal{L}_{\Bt} \GR \,=\, 0  ,
\label{eq:axifluid}
\eeq 
reduce the Euler Eq. (\ref{eq:reeq}) to the simple algebraic equation  
\beq
\frac{h}{u^t} = C,
\label{eq:ei}
\eeq
where $C$ a constant to be determined.
Note that $u^t$ is computed by the normalization condition $u^\GA u_\GA=-1$. In this work we also assume 
that neutron stars are ``cold'' i.e. they can be described by a zero-temperature equation of state (EOS), 
\beq
\GE=\GE(\GR),\quad  p=p(\GR),
\label{eq:barotropic} 
\eeq
or, equivalently,  $p=p(\GE)$. This kind of one-parameter EOS is called barotropic.
Equations (\ref{eq:psi})-(\ref{eq:hab}) under gauge conditions (\ref{eq:maximal_slicing}-\ref{eq:Dirac_gauge}) 
and Eq, (\ref{eq:ei}) determine all gravitational and fluid variables.

\subsubsection{Triaxial rotating neutron stars}
\label{sec:rnstri}

For nonaxisymmetric rotating neutron stars we impose time symmetry on the three metric 
but helical symmetry (stationarity in the rotating frame) on the extrinsic curvature
\beq
\mathcal{L}_{\Bt} \GG_{ij} =0, \qquad\mbox{and}\qquad \mathcal{L}_{\Bk} K_{ij} =0 .
\label{sec:tritime}
\eeq
Under conditions (\ref{sec:tritime}) we have 
$\tilde{u}_{ij}=0$ in Eq. (\ref{eq:MC}) and $\mathcal{L}_{\GA\Bn} (\cdots) = - \mathcal{L}_{\Bom} (\cdots)$
where $\GO^\GA = \GB^\GA + \Omega \GP^\GA$ the so-called corotating shift vector $\GO^\GA$.

The helical symmetry condition for the fluid variables 
\beq
\mathcal{L}_{\Bk} (h u_\GA) \,=\, \mathcal{L}_{\Bk} \GR \,=\, 0  ,
\label{eq:axifluid}
\eeq 
reduce the Euler Eq. (\ref{eq:reeq}) to Eq. (\ref{eq:ei}) similarly to the stationary and axisymmetric case.

\subsection{Binary neutron stars in the IWM formalism}
\label{sec:bns}

Binary neutron stars are computed within the IWM formalism where
\beq
h_{ij} = 0, \qquad\mbox{and}\qquad  K=0,
\label{eq:iwm}
\eeq
which means that Eq. (\ref{eq:hab}) is absent while in Eqs. (\ref{eq:psi})-(\ref{eq:lapse})
the magenta terms in the sources (\ref{eq:HC})-(\ref{eq:trdKijdt}) are zero. The spacetime helical symmetry 
\beq
\mathcal{L}_{\Bk} \GG_{ij} \,=\, \mathcal{L}_{\Bk} K_{ij} \,=\, 0 
\label{eq:binsym}
\eeq 
implies that Eq. (\ref{eq:taij}) can be written as 
\beq
\tA_{ij} = \frac{1}{2\GA} (\tilde{\mathbb L} \tilde{\GO})_{ij} 
\label{eq:bintAij}
\eeq
where $\GO^i$ the corotating shift.

The 4-velocity of the fluid here can be written as  
\beq
u^\GA := u^t (k^\GA + V^\GA), 
\label{eq:4vel}
\eeq
where $V^i$ is the fluid velocity in the corotating frame, 
and the helical Killing vector $k^\GA$ refers to a binary system
having orbital angular velocity $\Omega$.
For irrotational binaries \cite{Bonazzola97,Asada1998,Shibata98,Teukolsky98}, we have $\GO_{\GA\GB}=0$, so the specific
enthalpy current $hu_\GA$ can be derived from a potential $hu_\GA = \nabla_\GA
\Phi$. In order to allow for arbitrary spinning binary configurations, a
3-vector $s^i$ is introduced according to \cite{Tichy11}
\begin{equation}
\hat{u}_i := \GG_i^\GA hu_\GA = D_i\Phi + s_i\,,  \label{eq:uih_dec}
\end{equation}
where the $D_i\Phi$ part corresponds to the ``irrotational part'' of the flow
and $s^i$ the ``spinning part'' of the flow.  
Equations (\ref{eq:4vel}) and (\ref{eq:uih_dec}) yield
\begin{equation}
V^i=\frac{D^i\Phi+s^i}{hu^t}-\GO^i  . \label{eq:irspuih}
\end{equation}

For arbitrary spinning binaries  under the assumptions of helical symmetry 
\beq
\mathcal{L}_{\Bk}(h u_\GA) = 0    \label{eq:helsym}
\eeq
and the additional assumption for the spin of the neutron star
\beq
\mathcal{L}_{\boldsymbol V}(s_\GA) = 0 \ ,  \label{eq:assumspin}
\eeq
yields
\begin{equation}
\frac{h}{u^t} + V^j D_j\Phi = C\,, 
\label{eq:irspei}
\end{equation}
where $C$ is a constant to be determined. 

The conservation of rest mass and  Eq. (\ref{eq:irspuih}) for the fluid velocity
will produce an extra elliptic equation for the  fluid potential $\Phi$ 
\beqn
\nabla^2\Phi & = & -\frac{2}{\GC}\pd_i\GC\pd^i\Phi + \GC^4 \GO^i\pd_i(hu^t)  
+   [\GC^4hu^t \GO^i-\pd^i\Phi]\pd_i\ln\left(\frac{\GA\GR}{h}\right)  \nonumber \\
& \phantom{=} & -\GC^{4}\left[\pd_i s^i + s^i \pd_i\ln\left(\frac{\GA\GR\GC^{6}}{h}\right)\right]
:= \mathcal{S}_\Phi\,. \label{eq:cbm4}
\eeqn
Note that since $h_{ij}=0$, the conformal geometry is flat $\TDD{\GG}{i}{j}=f_{ij}$
and therefore $\pd_i\Phi=\TD{D}{i}\Phi$.
The boundary for the fluid is represented by the surface of the star;
hence the boundary condition for Eq. (\ref{eq:cbm4}) will be of
von Neumann type, that is, in terms of derivatives of the rest-mass
density and of $\Phi$
\begin{equation}
\left[\left(\GC^4hu^t \GO^i-\pd^i\Phi-\GC^{4} s^i\right)
       \pd_i\GR \right]_{{\rm surf.}} =0\,. 
\label{eq:bcphi}
\end{equation}
Equations (\ref{eq:psi})-(\ref{eq:lapse}) under maximal slicing gauge condition and 
Eqs. (\ref{eq:irspei}), (\ref{eq:cbm4}) determine all gravitational and fluid variables.
As in the rotating star case $u^t$ is determined from the normalization of the 4-velocity.
A barotropic EOS is assumed.

\section{Numerical methods}
\label{sec:nm}

All elliptic equations are solved using the
representation theorem of partial differential equations through a Picard type of iteration.  
Starting from 
\begin{equation}
\nabla^2 f = S\, 
\end{equation}
where $S$ is a nonlinear function of $f$ (here $f$ can be any of $\{\GC,\GA,\TD{\GB}{i},h_{ij},\Phi\}$ and 
$S$ any combination of them), 
and using an appropriate Green's function
\begin{equation}
\nabla^2 G(x,x') = -4\pi \GD(x-x')\,,   
\end{equation}
that satisfies certain boundary conditions, a solution for $f$ can be written as
\beq
 f(x) = -\frac{1}{4\pi}\int_{V} G(x,x')S(x') d^{3}x' 
+ \frac{1}{4\pi} \int_{\pd V} \left[G(x,x')\na'^{a} f(x') - f(x')\na'^a G(x,x') \right]dS'_a\,. 
\label{eq:GreenIde}
\eeq
where $V$ is the domain of integration, $x,x' \in V\subseteq\Sigma_t$,
the initial spacelike hypersurface. The volume $V$ and its boundary 
$\pd V$ depend on the coordinate system used and are different for isolated rotating neutron stars
and binary neutron stars. We will explain this difference in the next sections.
This method is widely known as the KEH method \cite{Komatsu89,Komatsu89b}.

For the evaluation of the integrals in Eq.(\ref{eq:GreenIde}), 
a multipole expansion of $G(x,x')$ in associated Legendre functions, $P_\ell^{~m}$,
on spherical coordinates is used; 
\beq
 G(x,x') := \sum_{\ell=0}^\infty g_\ell(r,r') \sum_{m=0}^\ell \epsilon_m \, \frac{(\ell-m)!}{(\ell+m)!}  
\times P_\ell^{~m}(\cos\theta)\,P_\ell^{~m}(\cos\theta') \cos [m(\phi-\phi')]\,, 
\label{eq:greenexp}
\eeq
where the radial Green's function $g_\ell(r,r')$ depends on boundary conditions 
and the coefficients $\epsilon_m$ are  $\epsilon_0 = 1$, and $\epsilon_m = 2$ for $m\ge 1$. 
In practice the sum over $\ell$ is truncated at $\ell=L$. The standard value used for single 
rotating stars as well as binaries is $L=12$. On the other hand highly rotating black holes and magnetized
neutron stars require a larger $L\lesssim 60$ .

\subsection{Single rotating neutron stars}
\label{sec:nmrns}

For a single rotating neutron star the only elliptic equations to be solved are
the gravitational Eqs. (\ref{eq:psi})-(\ref{eq:hab}). We employ a single spherical coordinate system that covers
the region $[r_a,r_b]\times[0,\pi]\times[0,2\pi]$, where $r_a=0$ and
$r_b\sim O(10^6 M)$,  $M$ being the total mass of the system. 
The boundary conditions (\ref{eq:bcpsal}) are applied at $r=r_b$.
Coordinate grids $(r_i,\theta_j,\phi_k)$ with 
$i = 0, \cdots, N_r$, $j = 0, \cdots, N_\theta$, 
and $k = 0, \cdots, N_\phi$, are freely specifiable except for the points at the boundary of the computational domain, 
The grid setup is the same as in \cite{Uryu2016a}: 
the radial grid intervals $\Delta r_i := r_i-r_{i-1}$ are 
constant when $r_i\in [r_a,r_c]$ $(i = 1, \cdots, \Nrm)$, which typically extends up to $\sim 1.25 R_e$, 
where $R_e$ is the equatorial radius of the star, and 
increase thereafter in a geometric progression with $\Delta r_i = k \Delta r_{i-1} $. Here 
$k$ is a constant, and  $r_i\in[r_c,r_b]$ $(i = \Nrm+1, \cdots, N_r)$.  
For the angular coordinate grids $(\theta_j,\phi_k)$, 
we choose equally spaced grids.  
Definitions of the parameters for the grid setups are 
listed in Table \ref{tab:rnsgrid_param}.  

When applying Eq. (\ref{eq:GreenIde}) we use the Green's function $G(x,x')=1/|x-x'|$.  
For the quadratures, a 2nd order midpoint 
rule is used in $r$ and $\phi$ integrations, and a 4th order midpoint rule 
is used for $\theta$ integration. 
We also use a 2nd order finite difference formula for the $r$, $\theta$ 
and $\phi$ derivatives evaluated at the mid points 
$(r_{i+\frac12},\theta_{j+\frac12},\phi_{k+\frac12})
=((r_{i}+r_{i+1})/2,(\theta_{i}+\theta_{i+1})/2,
(\phi_{i}+\phi_{i+1})/2)$, except for a 3rd order finite difference formula  used in the first radial derivative 
evaluated at the mid points.  
For derivatives at the grid points $(r_{i},\theta_{j},\phi_{k})$ a 4th order finite difference formula is used.  
Note that we use the midpoint rule for numerical quadrature formula, 
and hence compute the source terms, Eqs. (\ref{eq:HC})-(\ref{eq:tfdKijdt}), at the midpoints of the grids.  

\begin{table}
\begin{center}
\begin{tabular}{lll}
\hline
$r_{a}$ &:& Radial coordinate where the grid $r_i$ starts.                \\
$r_{b}$ &:& Radial coordinate where the grid $r_i$ ends.              \\
$r_{c}$ &:& Radial coordinate between $r_{a}$ and $r_{b}$ where the      \\
&\phantom{:}& grid changes from equidistant to non-equidistant.        \\
$N_{r}$ &:& Total number of intervals $\Delta r_i$ between $r_{a}$ and $r_{b}$. \\
$\Nrm$ &: & Number of intervals $\Delta r_i$ in $[r_a,r_{c}]$. \\
$\Nrf$ &:& Number of intervals $\Delta r_i$ in $[0,R(\theta,\phi)]$. \\
$N_{\theta}$ &:& Total number of intervals $\Delta \theta_i$ for $\theta\in[0,\pi]$. \\
$N_{\phi}$ &:& Total number of intervals $\Delta \phi_i$ for $\phi\in[0,2\pi]$. \\
$L$ &:& Number of multipole in the Legendre expansion. \\
\hline
\end{tabular}  
\caption{Summary of parameters used for single rotating star configurations.
$R(\theta,\phi)$ denotes the neutron star surface.}
\end{center}
\label{tab:rnsgrid_param}
\end{table}

A relaxation parameter $\GL$ is used when
updating a newly computed variable $f \in \{\GC, \GA, \TD{\GB}{i}, h_{ij}, \Phi, \GR\}$. 
If $f^{(n)}(x)$ is the value at the $n$-th iteration, and $\hat{f}(x)$ the current result of the Poisson 
solver, Eq, (\ref{eq:GreenIde}), then the $(n+1)$-th iteration value will be
\begin{equation} 
f^{(n+1)}(x)\,:=\,\GL \hat{f}(x)+(1-\GL)f^{(n)}(x)\,,
\label{eq:relax}
\end{equation}
where $0.1\leq \GL\leq 0.4$. Usually $\GL=0.4$ for all variables except for $\Phi$ where $\GL=0.1$.
The criterion used by the \cocal{} to stop the iteration for a variable $f$ is based on
the relative error between two successive iterations and given by
\begin{equation}     
\mathtt{E}^f(x) = 2\frac{|f^{(n)}(x) - f^{(n-1)}(x)|}{|f^{(n)}(x)| + |f^{(n-1)}(x)|} < 10^{-6}\,,
\label{eq:convcri} 
\end{equation}     
for all points of the grids, and all variables.

\subsection{Binary systems}
\label{sec:nmbns}

\begin{figure}
\begin{center}
\includegraphics[width=0.75\textwidth]{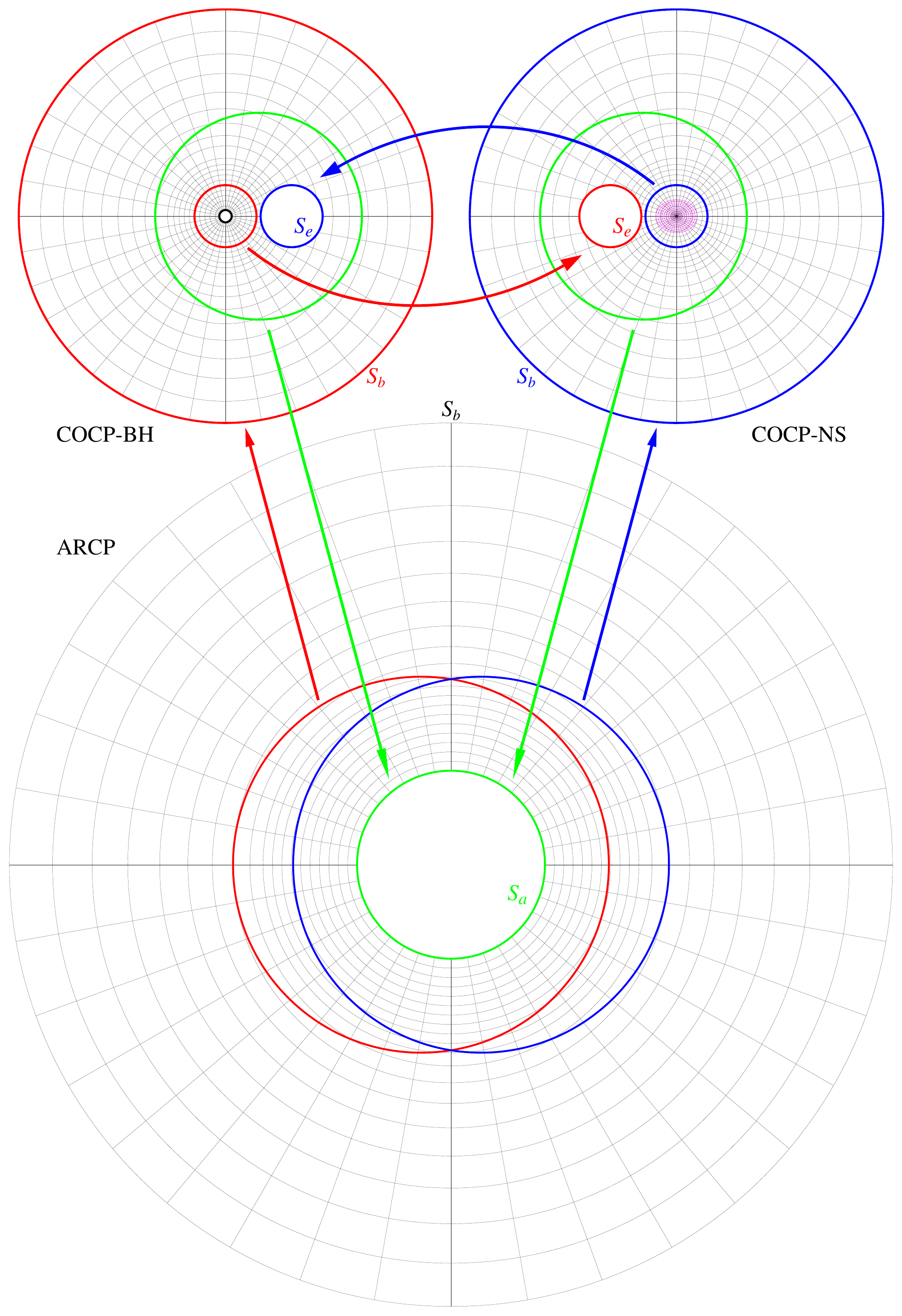}
\caption{A typical setup with multiple coordinate grid patches in the \cocal{} code 
for a black hole-neutron star system.  Left and right top patches are those 
for compact object coordinate patches (COCP) centered at the black hole and the neutron star correspondigly. 
The smallest circle with thick curve in COCP-BH is the sphere $S_a$ where 
the interior region is excised and certain boundary conditions are imposed.  
The ovals drawn in COCP-NS denote the neutron star.  
Bottom patch is that for asymptotic region coordinate patch (ARCP), centered 
at the mass center of the system.  
The arrows represent maps of potentials between the multiple patches.  
Note that the spheres $S_a$, $S_b$, and $S_e$ of these coordinate patches are 
distinct despite our use of a common symbol.  
The radius of each coordinate patch doesn't reflect the 
size used in actual computations.}
\label{fig:multipatches}
\end{center}
\end{figure}

For binary configurations the hypersurface $\Sigma_t$ is covered by at least
three coordinate systems as explained in detail in
\cite{Uryu2012,Tsokaros2012,Uryu:2012b,Tsokaros2015}. 
As an example of a general binary system, we show
in Fig. \ref{fig:multipatches} a black hole-neutron star 
($x-y$ cross section) which consists of the following: (i) Coordinate system
COCP-BH (top left coordinate system) centered around the black hole extending from an
inner sphere $S_a$ (small black circle at its center) of radius $r_a$ to a sphere $S_b$ of radius $r_b$ (red outer circle). The surface $S_a$ 
denotes the black hole region and is excised from the grid. If $M$ is
the mass of the system we have $r_a\sim O(M)$ while $r_b\sim O(100M)$. (ii)
Coordinate system COCP-NS (top right coordinate system) centered around the neutron
star extends from $r_a=0$ to a sphere $S_b$ of radius $r_b$ (blue outer circle).  Both COCP-BH
and COCP-NS contain an excised sphere $S_e$ (although we use the common symbol $S_e$ these spheres are
different in the two coordinate systems) which is introduced to improve the
angular resolution and reduce the number of multipoles {\it for resolving the
companion object in a given coordinate system}.  (iii) Coordinate system ARCP
that is positioned at the center of mass of the system and extends from an
inner sphere $S_a$ of radius $r_a\sim O(10M)$ (green inner circle) to an outer sphere $S_b$ of
radius $r_b\sim (10^6 M)$ (outer black circle).

In binary systems elliptic equations (\ref{eq:psi})-(\ref{eq:lapse}) are solved 
separately {\it in all three coordinate systems} (in Fig. \ref{fig:multipatches}) and a smooth solution emerges in an
iterative process. 
On the other hand Eq. (\ref{eq:cbm4}) is solved only in cordinate systems where a neutron star exists.
Therefore the computational cost of a binary system is significantly larger than that of 
a single rotating star. Convergence to a smooth solution is achieved because in any given coordinate system, in the
calculation of the surface integrals in Eq. (\ref{eq:GreenIde}) we use the values
of the potential functions $f$ from {\it another coordinate system} as
indicated by the red, green, and blue arrows in Fig. \ref{fig:multipatches}.
For example, when we calculate the contribution of the surface integral on the
red sphere $S_e$ inside the COCP-NS coordinate system (top right) we use the
values of the potentials on the corresponding red sphere from the COCP-BH
coordinate system (top left) as indicated by the red curved arrow. In this way
a given potential ($\GA,\GC,\GB^i$) from one coordinate system is
communicated to all others and at the end of the iterative process a smooth
solution is obtained in the whole computational domain.

\begin{table}
\begin{center}
\begin{tabular}{rll}
\hline
$r_a$: & Radial coordinate where the radial grids start. For      \\
\phantom{:}& the COCP-NS patch it is $r_a=0$. \\
$r_b$: & Radial coordinate where the radial grids end. \\
$r_c$: & Center of mass point. Excised sphere is located   \\
\phantom{:}& at $2r_c$ in the COCP patch. \\
$r_e$: & Radius of the excised sphere. Only in the COCP patch. \\
$N_{r}$: & Number of intervals $\Delta r_i$ in $[r_a,r_{b}]$. \\
$\Nrf$: & Number of intervals $\Delta r_i$ in $[0,R(\theta,\phi)]$ for the COCP-NS  \\
\phantom{:}& patch or in $[r_a,r_a+1]$ for the ARCP patch. \\
$\Nrm$: & Number of intervals $\Delta r_i$ in $[r_a,r_{c}]$. \\
$N_{\theta}$: & Number of intervals $\Delta \theta_j$ in $[0,\pi]$. \\
$N_{\phi}$: & Number of intervals $\Delta \phi_k$ in $[0,2\pi]$. \\
$L$: & Order of included multipoles. \\
\hline
\end{tabular}  
\caption{Summary of grid the parameters used for the binary systems and in particular the binary neutron stars
  computed here.  $R(\theta,\phi)$ denotes the neutron star surface. Every patch has its own set of parameters above.}
\end{center}
\label{tab:grid_param}
\end{table}

\section{Speedup and efficiency results}
\label{sec:nmbns}

The original serial COCAL consists of more than $400,000$ lines of Fortran 90 code and in order to use OpenMP
parallelization many of its loops had to be rewritten in a way that will efficiently utilizes the OpenMP capabilities. 
In particular, in multiple loops, the code inside the innermost one had to be written in a way that is {\it independent}
between threads. For example, 
in a typical triple loop over the $r$ (loop 1), $\GU$ (loop 2), and $\GP$ (loop 3) coordinates, 
any code between loop 1 and loop 2, and/or between loop 2 and loop 3, is written {\it only inside} 
the innermost loop 3.
The reason  is to utilize the {\tt collapse} clause which
collapses the multiple loops into a single one which is then divided among the multiple threads.
Most of our loops were three-dimensional but there were many that are four or even five-dimensional,
which result to large speedups through OpenMP parallelization.
For example, in a binary system Fig. \ref{fig:multipatches}, the 
calculation of the surface integral on the excised sphere $S_e$ uses the Legendre functions with respect
to the coordinate system at the center of $S_e$. Assuming $\theta_e$ is the spherical angle with respect to the z-axis
at the center of $S_e$, the term $\cos(\theta_e)$ will depend on all coordinates $r,\ \GU,\ \GP$ of that particular 
patch and therefore these functions become five dimensional, since they depend on the Legendre indices $\ell,\ m$ too.

Another commonly used command is the {\tt reduction} clause which is used to perform summations,
as for example in the calculation of the volume and surface integrals. In this case the code is written
in a way that the summation appears {\it only inside the innermost loop} so that a combination of the {\tt collapse} 
and the {\tt reduction} clause can execute this operation in multiple threads.
The {\tt reduction} clause is also used in finding the maximum
error in Eq. (\ref{eq:convcri}), for every variable in each coordinate system.
The {\tt private} clause is used extensively for local nested-loop variables so that multiple threads 
executing a parallel region are independently calculated.
Listing a variable as {\tt private} causes each thread that executes that construct to receive 
a new temporary variable of the same type and the multiple loop can be performed independently in parallel.

Below we will describe the two main modules of the PCOCAL code: a) the parallelized 
single rotating star module (Sec. \ref{sec:rns}), and b) the parallelized binary neutron star module (Sec. \ref{sec:bns}).
For the single star solver (a), the computational infrastructure is straightforward (one spherical grid), but the mathematical formalism 
used requires the solution of 14 elliptic equations. On the other hand for the binary solver (b) a simpler mathematical formalism
requires the solution of 6 elliptic equations in a more  intricate computational domain which involves at least three coordinate 
systems (see Fig. \ref{fig:multipatches}).

In order to quantify the speedup and efficiency of PCOCAL we define the following measures:
\begin{itemize}
\item $T_p(n)$: CPU wall-clock time of parallelized part of the code using $n$ threads.
\item $T_s$: CPU wall-clock time of serial (nonparallelized) part of the code. 
Input/Output (IO) and copying between arrays are excluded.
\item $T_{\rm IO}$: CPU wall-clock time of serial IO.
\item $T_m$: CPU wall-clock time of memory copy between arrays.
\end{itemize}
The reason for timing $T_m$ is because in COCAL/PCOCAL every main variable (e.g. the 3d conformal factor $\GC$)
exists in all coordinate systems (which are at least three) and is stored in a higher dimensional array (for the
case of $\GC$, in a 4d array) where the extra index determines the patch (COCP-1, COCP-2, or ARCP). When calculating
quantities in one patch and we want to access a quantity in another patch, an array copy is necessary. The time
for these copies is measured by $T_m$. 

The total time for a calculation is $T = T_p + T_s + T_{\rm IO} + T_m$.
We define the speedup of the code using $n$ threads as
\beq
S_p(n) = \frac{T_p(1)}{T_p(n)} ,\qquad S(n) = \frac{T(1)}{T(n)}
\label{eq:speedup}
\eeq
while the efficiency is defined as 
\beq
E_p(n) = \frac{S_p(n)}{n} ,\qquad E(n) = \frac{S(n)}{n}  .
\label{eq:effic}
\eeq
%
%
%
%
%
%

The speedup $S_p(n)$ and efficiency $E_p(n)$ refer to the parallelized part of code (which
constitutes the large majority of the solver), while  $S(n)$ and $E(n)$ refer to the total speedup and efficiency
which include the serial parts of the code, the input/output, and memory handling. Note that for the tests performed 
in this paper the input/output routines were neither optimized nor minimized (for diagnostic purposes) and thus
the total speedup shown underestimates the real one.

To make sure that the parallelized PCOCAL code produces the same results as COCAL itself, we perform a pointwise
check for \textit{every} parallelized subroutine against the serial code and confirm that the differences between
the two codes are $\lesssim 10^{-12}$ in all variables.

Most of the runs have been performed in  server A which has 36 cores with 2 threads per core (72 threads total)
and an Intel(R) Xeon(R) Gold 6254 CPU \@ 3.10GHz.
We have also used  server B which has 40 cores with 2 threads per core (80 threads total)
and an Intel(R) Xeon(R) Gold 6242R CPU \@ 3.10GHz.
Intel compiler (2021.3.0 20210609) is being used.




\subsection{Single rotating neutron star}
\label{sec:prns}

Following the methods of Sec. \ref{sec:rns} and Sec. \ref{sec:nmrns} we test our new code in three grid resolutions as can
be seen in Table \ref{tab:rnsreso}. Resolution H2 has $N_r\times N_\theta\times N_\phi=442368$ intervals (number of points 
in the $r,\GU,\GP$ directions are $N_r+1$, $N_\GU+1$, and $N_\GP+1$ respectively) while resolutions H3, H4
have twice and four times as many intervals in every dimension, i.e. 8 and 64 times more intervals in total respectively.
The default value of terms in the Legendre expansions is $L=12$. Below we also investigate the performance of the code
with respect to $L$.

\begin{table}
\begin{center}
\begin{tabular}{ccccccccc}
\hline
\hline
Type & $r_a$ & $r_b$ & $r_c$ & $N_r$ & $\Nrm$ & $\Nrf$ & $N_\theta$ & $N_\phi$  \\
\hline
H2 & 0 & $10^6$ & 1.25 & 192 &  80 &  64 &  48 &  48  \\
H3 & 0 & $10^6$ & 1.25 & 384 & 160 & 128 &  96 &  96  \\
H4 & 0 & $10^6$ & 1.25 & 768 & 320 & 256 & 192 & 192 \\
\hline
\hline
\end{tabular}
\caption{Three different resolutions used for the single rotating star tests.
  Parameters are shown in Table \ref{tab:rnsgrid_param}. The number of points that
  covers the largest star radius is $\Nrf$.
  Resolutions H2, H3, and H4 use 2, 10.3, and 80.1 GB of RAM correspondingly.
}
\end{center}
\label{tab:rnsreso}
\end{table}

\begin{algorithm}
\caption{Rotating star in the waveless formalism}
\begin{algorithmic}[1]
\Procedure{RNS}{}                              
    \State Interpolate variables to SFC        \Comment{$\color{green1}(8\%)\color{black}$}  
    \State Compute $K_{ij}\ \color{green1}(8\%)\color{black}$,            
           $C^i_{\ kj}\ \color{green1}(8\%)\color{black}$,                
           $\tRs_{ij}\ \color{green1}(11\%)\color{black}$                 
                                               \Comment{$\color{green1}(27\%)\color{black}$}
    \State Compute volume sources $\mathcal{S}_{\rm H}$,  Eq. (\ref{eq:HC})  
                                               \Comment{$\color{green1}(5\%)\color{black}$}  
    \State Compute volume sources $\mathcal{S}_{\rm tr}$,  Eq. (\ref{eq:trdKijdt}) 
                                               \Comment{$\color{green1}(6\%)\color{black}$} 
    \State Compute volume sources $\mathcal{S}_i$, Eq. (\ref{eq:MC})  
                                               \Comment{$\color{green1}(15\%)\color{black}$}  
    \State Compute volume sources ${\cal S}_{ij}$, Eq. (\ref{eq:tfdKijdt})
                                               \Comment{$\color{green1}(13\%)\color{black}$}   
    \State Compute right-hand side of Eq. (\ref{eq:gauge}) 
                                               \Comment{$\color{green1}(3\%)\color{black}$}    
    \State Compute $\GC,\GA,\TD{\GB}{i},h_{ij},\xi^i$, Eq. (\ref{eq:GreenIde}) 
                                               \Comment{$\color{green1}(11\%)\color{black}$}   
    \State Variable update, Eq. (\ref{eq:relax})   
                                               \Comment{$\color{green1}(2\%)\color{black}$} 
    \State Use Eq. (\ref{eq:ei}) to compute the rest-mass density $\GR$.
    \State Compute $\mathtt{E}^\GC,\mathtt{E}^\GA,\mathtt{E}^\GB_{i},\mathtt{E}^h_{ij},\mathtt{E}^\GR$ Eq. (\ref{eq:convcri})   
                                               \Comment{$\color{green1}(1\%)\color{black}$}  
    \If{$(\mathtt{E}^\GC,\ \mathtt{E}^\GA,\ \mathtt{E}^\GB_{i},\ \mathtt{E}^h_{ij},\ \mathtt{E}^\GR\ <\ 10^{-6})$}
        \State exit
    \EndIf    
\EndProcedure
\end{algorithmic}
\end{algorithm}

\begin{figure*}[h]
\includegraphics[width=0.49\textwidth]{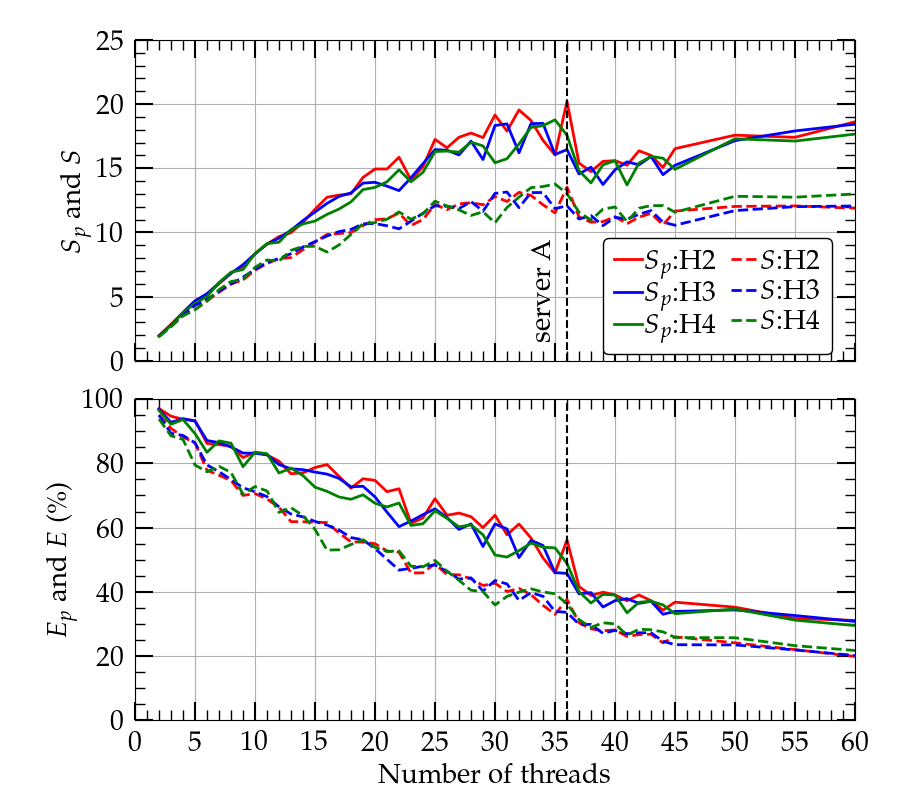}
\includegraphics[width=0.49\textwidth]{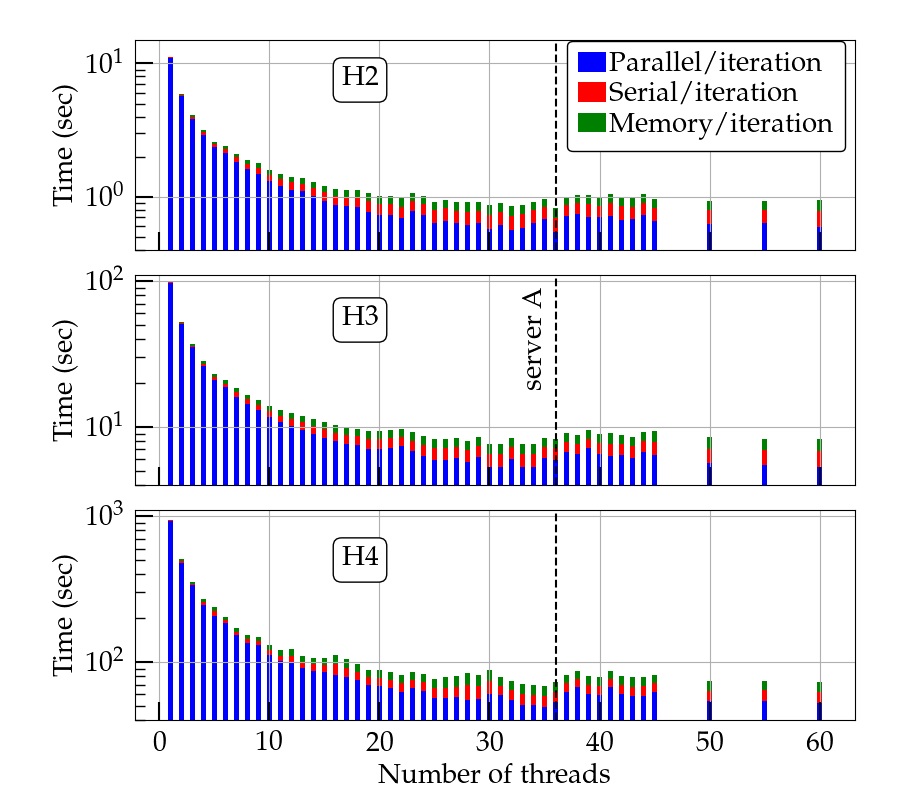}
\caption{Left panels: Speedup (top) and efficiency (bottom) of the parallelized (solid lines) and 
total code (dashed lines) of a single rotating star module in the 
waveless formalism for three different resolutions H2, H3, and H4. In all cases the maximum number of terms in the 
Legendre expansion is $L=12$. The vertical dashed line denotes the number of cores/CPU of server A.
Right panels: Performance time of the whole code in the three resolutions.}
\label{fig:rnsH234}
\end{figure*}

\begin{figure}[h]
\centering
\includegraphics[width=0.49\textwidth]{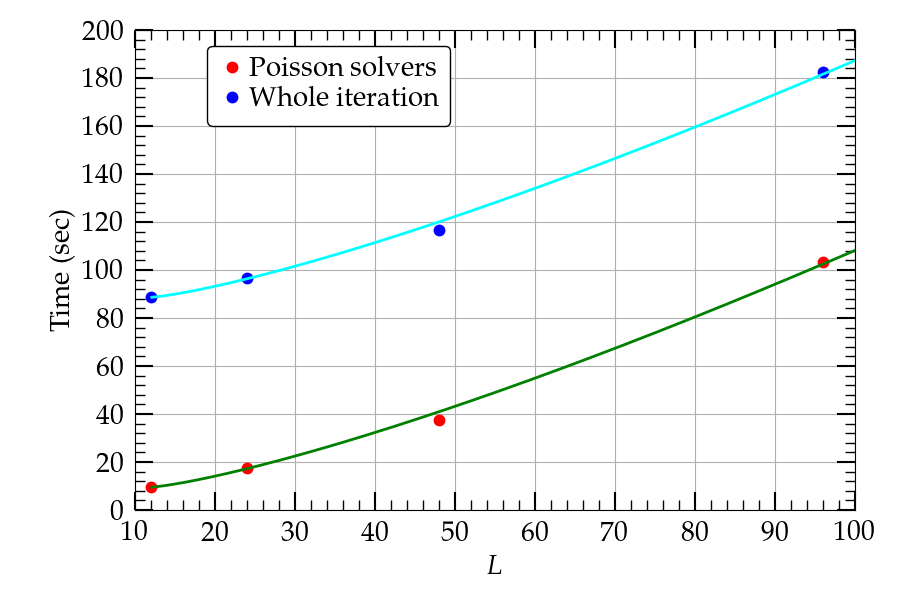}
\caption{Time for the Poisson solvers (red dots) as well as for a single iteration (blue dots), 
for the H3 resolution in a \textit{single core}, for various values of Legendre terms $L$. Solids lines represent fitting functions.}
\label{fig:rnsL}
\end{figure}

In Algorithm 1 we sketch the most salient steps taken for a solution. In parentheses with green fonts we show the 
percentage of time needed for a given calculation per iteration using a single core using the H3 resolution. 
Lines that do not 
have a number indicate that the time it took for completion was less than $1\%$. 
The rest of the time is spent in the calculation of diagnostics as well as further output.
Steps 2-12 are repeated iteratively until condition 13 is
fulfilled, and hence convergence to a solution is achieved. 
As we can see, the most time consuming routines are the computation of the momentum constraint
sources $\mathcal{S}_i$ (15\%), the calculation of the sources $\mathcal{S}_{ij}$  (13\%) followed by
the calculation of the conformal Ricci tensor $\tRs_{ij}$ (11\%), and the Poisson solvers themselves (11\%).
The reason that the three source arrays $\mathcal{S}_i$ of the momentum constraint took more time to
be computed than the six source arrays $\mathcal{S}_{ij}$, was mainly due to 
modular way COCAL implements various mathematical formulations. In particular for the waveless formulation
the sources (right hand side of the Poisson equations) are split into two parts: i) the part that comes from
the conformal flat part of the metric, and ii) the ones that come from the nonconformal one.
In this way, one has the flexibility of choosing a specific method (conformal flat vs nonconformal flat) 
without repeating the writing of a code.
The disadvantage is that triple loops all over the gridpoints may be repeated, therefore speed is sacrificed
in view of modularity. In the computation of the momentum constraint sources $\mathcal{S}_i$ for the waveless
formulation (which is not conformally flat) both the fluid part and the gravitational part of the source
computation are done twice, which leads to a larger time footprint than the six source arrays $\mathcal{S}_{ij}$.
Overall the bottleneck for the rotating neutron star module in the waveless formulation is the computation of the sources
(lines 2-8) rather than the Poisson solvers (line 9). As we will see this is in contrast with the binary neutron
star module in the conformal flat approximation, where the latter dominate over the former.

\begin{figure*}[h]
\centering
\includegraphics[width=0.49\textwidth]{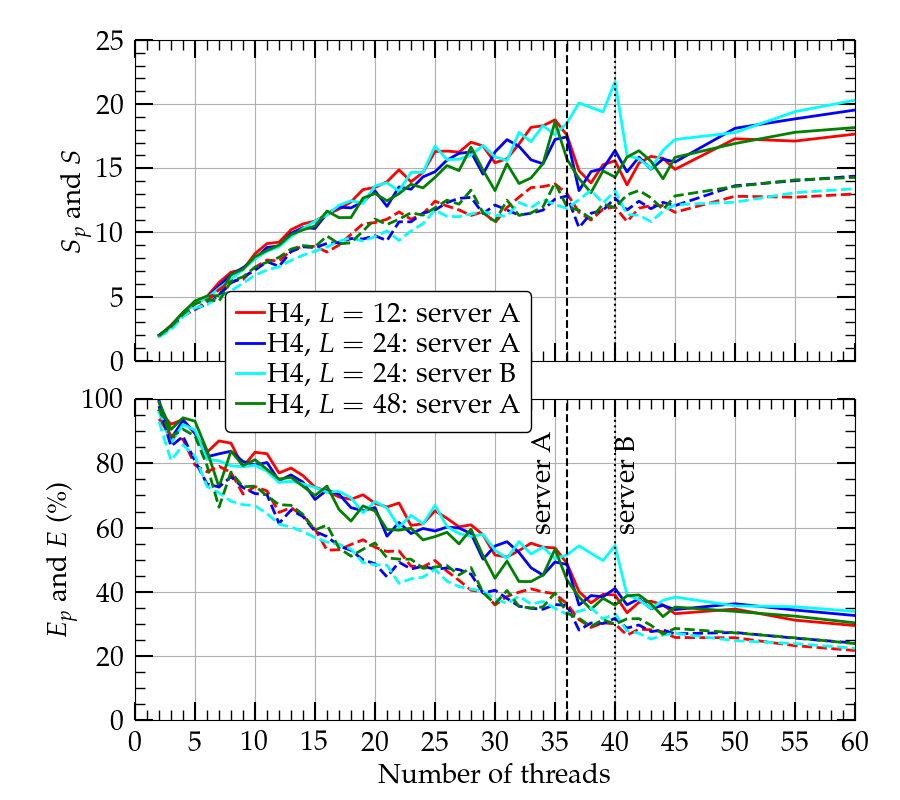}
\includegraphics[width=0.49\textwidth]{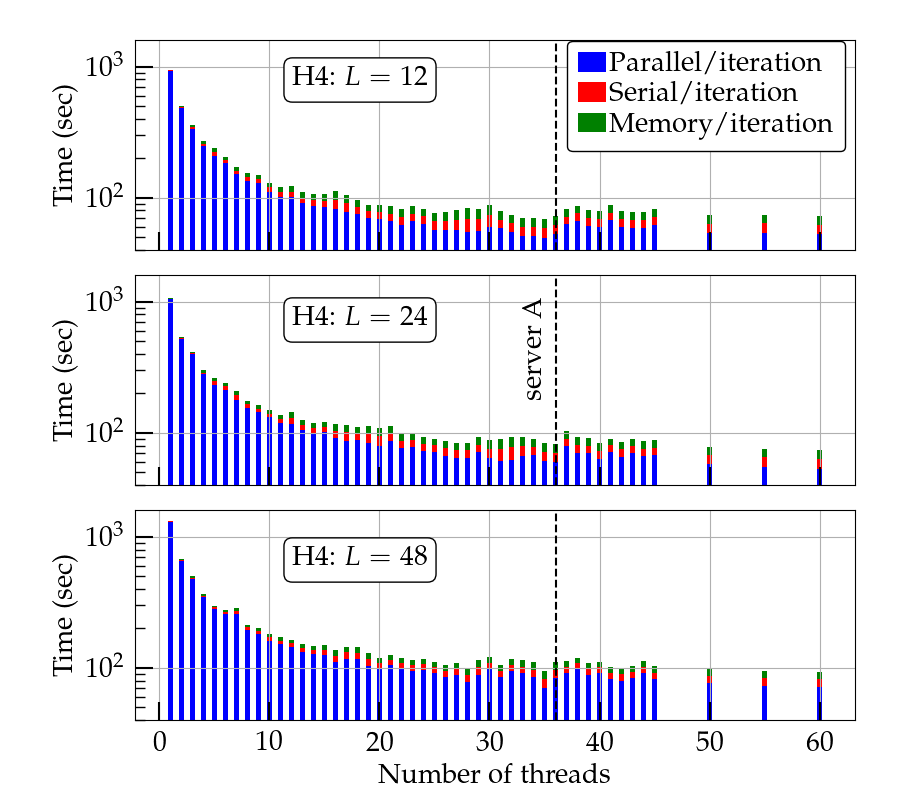}
\caption{Same as in Fig. \ref{fig:rnsH234} but only for resolution H4 with $L=12,\ 24,\ 48$. Speedup and efficiency 
in a second (server B) is also shown for $L=24$ (cyan curve on the left panels).
As in Fig. \ref{fig:rnsH234}, solid lines in the left panels correspond to $S_p$ and $E_p$, while dashed lines to 
$S$ and $E$. Vertical dashed lines denote the number of cores/CPU of servers A and B.}
\label{fig:rnsH4L}
\end{figure*}

In Fig. \ref{fig:rnsH234} on the left panels we plot the speedup $S_p$ (top) and efficiency $E_p$ (bottom) of the parallelized part
of the code (solid lines),
as well as the corresponding measures $S$ and $E$ for the total code (dashed lines).
As we mentioned above, $S$ underestimates the real speedup since the input/output routines were neither optimized nor 
minimized (for diagnostic purposes). Thus in the discussion below we focus on the speedup $S_p$ and efficiency $E_p$.
One characteristic of the speedup is that for all resolutions it reaches a maximum at $\sim 36$ threads, drops slightly
afterwards, from which point it continues to increase. The maximum speedup of the parallelized part of the code 
is $18-20$ times the serial one (when $\lesssim 60$ threads are used) which can be
achieved with a minimum of $\sim 36$ threads. At that point the efficiency is $\sim 50\%$.
At maximum speedup the whole rotating neutron star code 
(including the serial routines that are nonparallelized, such as the calculation of the parameters 
$\Omega$ in Eq. (\ref{eq:4velrns}), $C$ in Eq. (\ref{eq:ei}), and the coordinate scaling $R_0$ \cite{Uryu2016a,Tsokaros2015}) 
is $\sim 12$, $13$, and $14$ times faster than the 
serial analogue for resolutions H2, H3, and H4 and occurs at $36$, $34$, and $35$ threads respectively.
This can be seen in Fig. \ref{fig:rnsH234} on the right panels, where blue bars signify the parallelized routines
of the code, red bars the serial part of the code, and green bars memory copy between arrays.
From this plot we see that the parallelized code constitutes the vast majority of the total number of subroutines
and is responsible for the achieved speedup. Also we observe that the highest the resolution (H4), the larger the
total speedup, which is a promising result for high resolution campaigns.

In all tests presented in Fig. \ref{fig:rnsH234} we used $L=12$ terms in the Legendre expansion, Eq. (\ref{eq:greenexp}), 
and therefore all the integrals, Eqs. (\ref{eq:GreenIde}). 
It is experimentally found that such number of terms leads to accurate results without compromising the speed
of the code. $L=12$ is commonly used in all nonmagnetized rotating star calculations 
\cite{Huang08,Uryu2016a,Uryu2016b,Uryu2017,Zhou2017xhf,Zhou2019hyy} as well as
in the binary neutron star calculations \cite{Tsokaros2015,Tsokaros2018} that we will mention in the next section.
On the other hand accurate magnetized rotating neutron stars or black hole-disk solutions require a larger number
($\lesssim 60$) of terms to be included in the Legendre expansions.
In Fig. \ref{fig:rnsL} we plot the time of a single iteration (blue dots), 
as well as the time for the Poisson solvers alone (red dots), using the H3 resolution and a \textit{single core}. 
Increasing the number of Legendre terms $L$, increases the time spent on the Poisson solvers as
\beq
\mbox{Time} \approx 0.32(L-12)^{1.28} + c
\label{eq:Lscale}
\eeq
where $c=9.52$. This fitting function is plotted by a solid green line. At the same time, the time spent 
on the whole iteration is also fitted by the same relation Eq. (\ref{eq:Lscale}), but $c=88.63$\footnote{
As reported in Algorithm 1, line 9, the percentage of time spent for the Poisson solvers
with respect to the whole iteration, when $L=12$, is approximately $11\%$.}, and is 
plotted with a cyan curve. In other words, for a given resolution, the increase of the number of Legendre
terms affects essentially the Poisson solvers as well as the whole iteration in the same manner and scale
as $\sim L^{1.28}$.

In order to evaluate our new code with respect to the number $L$ we
performed the same tests using the H4 resolution for three different number of Legendre terms $L=12,\ 24,\ 48$. The results
are shown in the left panels of Fig. (\ref{fig:rnsH4L}) with red, blue and green curves in server A.
In addition we show with a cyan curve the speedup and efficiency of the $L=24$ run in server B.
The first conclusion is that the speedup is approximately preserved when the number of Legendre terms is increased. All
runs in server A have a maximum at $\sim 35$ threads. Beyond that point, and similarly to Fig. \ref{fig:rnsH234}, 
the speedup drops and then starts to increase again. This behavior is qualitatively the same as the run with $L=24$
at server B with the maximum now happening at $40$ threads and a speedup larger than $20$. Given than
the server A CPUs have 36 cores, while the server B CPUs have 40, we conclude that peak speedup with minimum amount of
threads happens approximately at the number of cores a CPU has. 
The larger the number of cores the larger the speedup will be. On the 
right panels we plot the total performance time for the $L=12,\ 24,\ 48$ in the H4 resolution and server A.
We find that the maximum speedup of the total rotating neutron star code (including the routines that are nonparallelized)
is $\sim 14$, $13$, and $14$ times faster than the
serial analogue in the H4 resolutions for $L=12,\ 24,\ 48$,  and occurs at $35$, $36$, and $35$ threads respectively.
Therefore the speedup is preserved when the number of Legendre terms is increased for the whole code as well, which is
expected, given the fact that the parallelization scheme covers all the main components of PCOCAL.

\begin{table*}[h]
\centering
\begin{tabular}{cl|ccccccccccccc}
\hline
\hline
Type & Patch  & $\ r_a\ $ & $\ r_b\ $ & $\ r_c\ $ & $\ r_e\ $ & 
$\ \Nrf\ $ & $\ \Nrm\ $ & $\ N_r\ $ & $\ N_\theta\ $ & $\ N_\phi\ $ & $\ L\ $  \\
\hline
E2.5  & ${\rm COCP}$ & $0.0$ & $40$   & $2.51$ & $2.375$ & $80$  & $151$ & $288$ & $72$  & $72$ & $12$  \\ 
     & ${\rm ARCP}$ & $5.0$ & $10^6$ & $6.25$ & $-$     & $24$  & $30$  & $288$ & $72$  & $72$ & $12$  \\
\hline
E3.0  & ${\rm COCP}$ & $0.0$ & $40$   & $2.51$ & $2.35$  & $100$ & $188$ & $384$ & $96$  & $96$ & $12$ \\ 
     & ${\rm ARCP}$ & $5.0$ & $10^6$ & $6.25$ & $-$     & $32$  & $40$  & $384$ & $96$  & $96$ & $12$  \\
\hline
E3.5  & ${\rm COCP}$ & $0.0$ & $40$   & $2.51$ & $2.355$ & $152$ & $286$ & $576$ & $144$  & $144$ & $12$  \\
     & ${\rm ARCP}$ & $5.0$ & $10^6$ & $6.25$ & $-$     & $48$  & $60$  & $576$ & $144$  & $144$ & $12$ \\
\hline
\hline
\end{tabular}
\caption{Three different grid structure parameters used for the circular
  binary computation in COCAL. 
  The amount of RAM used in each resolution is written in the first column.
  All variables are explained in
  Table~\ref{tab:grid_param} and the distances are in normalized
  quantities \cite{Tsokaros2015}. There are two COCP grids centered around each neutron star, and one ARCP
  centered at the center of mass of the system (see Fig. \ref{fig:multipatches}).
  Resolutions E2.5, E3.0, and E3.5 use 5.0, 9.9, and 30.6 GB of RAM correspondingly.
}
\label{tab:bnsgrids}
\end{table*}

\subsection{Binary neutron stars }
\label{sec:pbns}

Following the methods of Sec. \ref{sec:bns} and Sec. \ref{sec:nmbns} we test our new code in three resolutions as
in Table \ref{tab:bnsgrids}. Resolution E2.5 has $N_r\times N_\theta\times N_\phi=1492992$ intervals in each of the 
three coordinates systems (two COCPs and one ARCP in Fig. \ref{fig:multipatches}) while resolutions E3.0 and E3.5
have $\sim 1.3$, and $2$  times as many intervals in every dimension, i.e. $2.4$ and $8$ times more intervals in total respectively.
In all binary cases we use $L=12$ number of terms in the Legendre expansions.
Notice that the BH coordinate system (COCP-BH, top left, red patch in Fig. \ref{fig:multipatches}) is now replaced by a NS
coordinate system, similar to the blue, top right patch COCP-NS. Therefore there is no inner surface $S_a$ (no boundary conditions), 
and $r_a=0$ for the red patch in Fig. \ref{fig:multipatches} for the binary neutron/quark case.

\begin{algorithm}
\caption{Binary neutron star }
\begin{algorithmic}[1]
\Procedure{NSNS}{}       
    \ForAll{Coordinate Systems $A$}
      \State Interpolate variables from SFC  \Comment{$\color{green1}(3\%)\color{black}$}  
      \State Compute $K_{ij}$                \Comment{$\color{green1}(9\%)\color{black}$}  
      \State Compute volume source Eq. (\ref{eq:HC}), $\mathcal{S}_{\rm H}$
      \State Compute volume source Eq. (\ref{eq:trdKijdt}),  $\mathcal{S}_{\rm tr}$
      \State Compute volume source Eq. (\ref{eq:MC}), $\mathcal{S}_i$
                                             \Comment{$\color{green1}(6\%)\color{black}$}
      \State Compute sources on excised surface $S_e$ 
                                             \Comment{$\color{green1}(1\%)\color{black}$}   
      \State Compute $\GC,\GA,\TD{\GB}{i}$ Eq. (\ref{eq:GreenIde})
                                             \Comment{$\color{green1}(41\%)\color{black}$}  
      \State Variable update Eq. (\ref{eq:relax})
                                             \Comment{$\color{green1}(1\%)\color{black}$}
      \State Use Eq. (\ref{eq:irspei}) to compute the rest-mass density $\GR$.
      \State Compute $\Phi$ Eq. (\ref{eq:cbm4}), (\ref{eq:GreenIde})
                                             \Comment{$\color{green1}(4\%)\color{black}$}   
      \State Compute $\mathtt{E}^\GC(A),\mathtt{E}^\GA(A),\mathtt{E}^\GB_{i}(A),\mathtt{E}^\GR(A), \mathtt{E}^\Phi(A)$ 
                                             \Comment{$\color{green1}(1\%)\color{black}$}   
      \State Variable copy between CSs
                                             \Comment{$\color{green1}(2\%)\color{black}$}   
    \EndFor
    \If{$(\mathtt{E}^\GC(A),\ \mathtt{E}^\GA(A),\ 
          \mathtt{E}^\GB_{i}(A),\ \mathtt{E}^\GR(A),\ \mathtt{E}^\Phi(A)\ <\ 10^{-6}$ \newline \hspace*{25pt} for all $A$)}
        \State exit
    \EndIf    
\EndProcedure
\end{algorithmic}
\end{algorithm}

\begin{figure*}[h]
\centering
\includegraphics[width=0.49\textwidth]{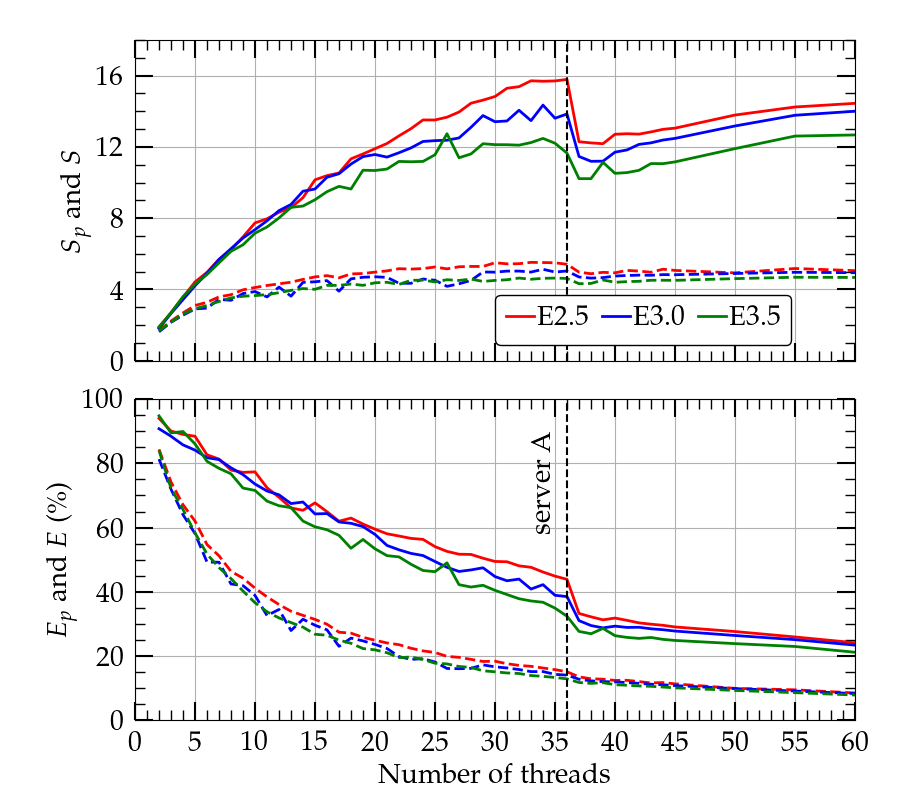}
\includegraphics[width=0.49\textwidth]{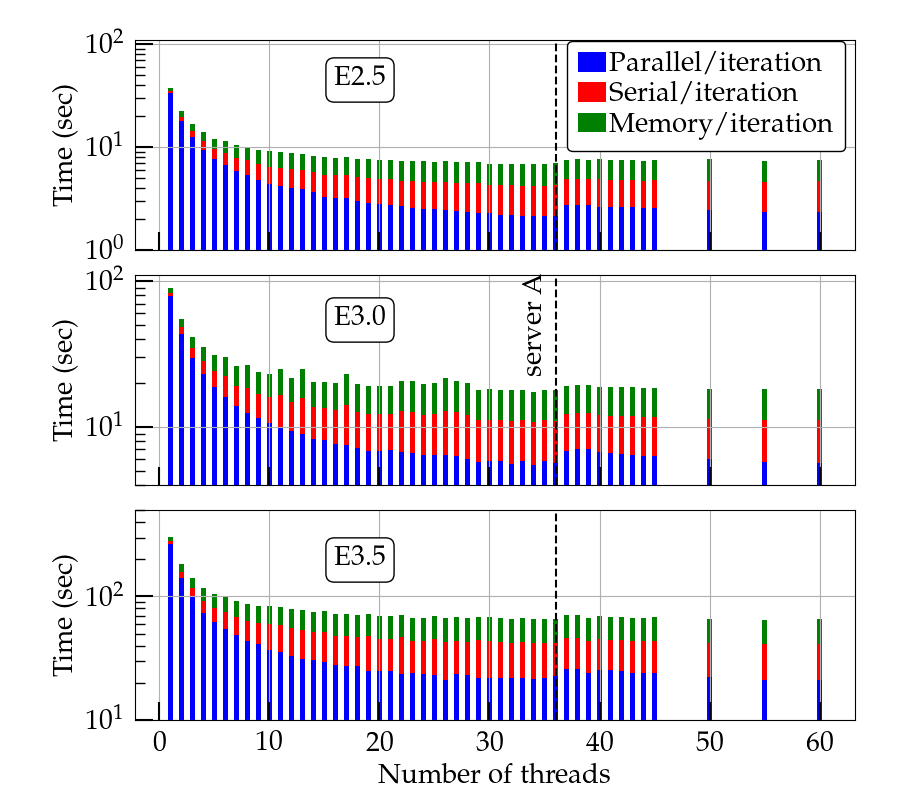}
\caption{Same as in Fig. \ref{fig:rnsH234} but for binary neutron stars. The resolutions used are shown in Table
\ref{tab:bnsgrids} and are E2.5, E3.0, and E3.5.
As in Fig. \ref{fig:rnsH234}, solid lines in the left panels correspond to $S_p$ and $E_p$, while dashed lines to 
$S$ and $E$.}
\label{fig:bns}
\end{figure*}

In Algorithm 2 we sketch the most important steps taken for a binary neutron star solution using a single core in
the E3.0 resolution. 
As in Algorithm 1 we report the percentage of time 
needed for the completion of each step with green fonts inside a parenthesis, while when such number is absent it means 
that the time it took for completion was less than $1\%$. 
The rest of the time is spent in the calculation of diagnostics as well as further output.
Steps 2-14 are repeated iteratively until condition in line 15 is
fulfilled. The main differences between the RNS and NSNS modules are: i) Steps 3-15 are performed in three
coordinate systems (two for COCP-NS and the ARCP) as seen in Fig. \ref{fig:multipatches} instead of one in the RNS module. 
ii) The calculation of every potential  $\GC,\GA,\TD{\GB}{i}$ in the two COCP-NS coordinate systems is significantly more 
involved because of the excised sphere $S_e$\footnote{The reason for the existence of $S_e$ in the COCP patches
is explained in detail in \cite{Uryu2012,Tsokaros2012,Uryu:2012b}.}.
iii) The additional calculation of the velocity potential $\Phi$ ($\sim 4\%$) (absent in the RNS module) 
is another important difference between the NSNS module and the RNS one.
iv) Since for binary neutron stars we solve for conformally flat initial data, potetials
$h_{ij}=0$ and in addition the source terms 
(Eqs. (\ref{eq:HC}), (\ref{eq:MC}), (\ref{eq:trdKijdt})) are significantly simpler since the magenta terms are zero.

As mentioned above, surface integrals (see Eq. (\ref{eq:GreenIde})) on $S_e$ 
do not exist in the single coordinate system of the RNS module. When computing those integrals as well as the surface
integrals at $S_b$ and $S_a$ in a given coordinate system (e.g. on COCP-NS-1) the integrands are computed using the variables
of another coordinate system (e.g. from COCP-NS-2). In this way solutions between coordinate systems communicate between each other
in order to achieve a smooth solution everywhere. Therefore to compute the integrands on the surface integrals
at $S_a$, $S_b$, and $S_e$ three-dimensional interpolations from nearby points of another coordinate system is
needed. In total the Poisson solvers take $\sim 41\%$ of the time of an iteration. The time of the Poisson solvers
in each COCP-NS patch is $\sim 17\%$ while in the ARCP is $\sim 7\%$,  
The big difference between them is
due to the fact that patch ARCP does not have an excised sphere $S_e$. The surface 
integral in the off-center sphere $S_e$ is the most time consuming operation ($\sim 11\%$),
and it involves the computation of the corresponding associated Legendre 
functions which are five-dimensional arrays of ($r,\GU,\GP,\ell,m$). In the current implementation this time consuming
operation is done on the fly every time such an integral is calculated in order to have a smaller memory footprint. Note
that these functions do not change during the iteration procedure and in principle they can only be calculated 
once. Such an array in the E3.5 resolution will be $\sim 9$ GB. 


In Fig. \ref{fig:bns} 
on the left panels we plot the speedup $S_p$ (top) and efficiency $E_p$ (bottom) of the parallelized part of the
binary NSNS code 
(solid lines), as well as the corresponding measures $S$ and $E$ for the total code (dashed lines).
Similar to the RNS module, Fig. \ref{fig:rnsH234}, we find that for all resolutions the speedup increases
until a certain number of threads, then it exhibits a sudden small drop, from which point it continues to increase
with the number of threads. 
One difference with respect to Fig. \ref{fig:rnsH234} is that the speedup curves are distinct for 
resolutions E2.5, E3.0, and E3.0, while for the RNS module we observe a broad overlap of these curves between resolutions
H2, H3, H4. 
The reason for this behavior is the communication between the different coordinate systems in the binary code
which is absent in the single neutron star module. 
Despite of that the speedup of the parallelized code reaches values
of $12-16$ using just $\sim 36$ threads. The efficiency at that  point is $\sim 40\%$. 
This speedup can further increase if more than 60 threads are used. 
Another finding in Fig. \ref{fig:bns} 
is the fact that for the highest resolution E3.5 the speedup is broadly constant from $\sim 26$ to $\sim 35$ threads
which is due to the domination of data communication between the different grids in the calculation process.
On the left panels of Fig. \ref{fig:bns}
we also see that the total binary code (including the routines that are nonparallelized) 
is $6$, $5$, and $5$ times faster than the 
serial analogue for resolutions E2.5, E3.0, and E3.5 and occurs at $33$, $34$, and $26$ threads respectively. 
The distribution of times in the whole code can be seen on the right panels of Fig. \ref{fig:bns}.
In the future
we will address the data communication and memory hadling in order to make the binary speedup to scale 
similarly to the RNS code. In practice a binary neutron star in quasicircular orbit that took approximately six
days of continuous computation in the E3.0 resolution with the COCAL code, with PCOCAL it needs approximately
one day using a modest amount of $\sim 30$ threads.

The differences between the PCOCAL and COCAL for a binary neutron or quark star are $\lesssim 10^{-12}$
in all coordinate systems for the gravitational variables $\GC, \GA, \GB^i$ and fluid variables $\GR, \Phi$ 
as well as for the constants $\Omega, C$ and $R_0$ that appear in the calculation.

\section{Spinning binary quark stars}
\label{sec:sbqs}

Given the fact that the state of matter at supranuclear densities is an open question, there exists the possibility
that conventional neutron stars contain in their core a phase transition to strange-quark matter i.e. up, down, and 
strange free quarks. 
A quark-gluon plasma has been observed in heavy ion collision at CERN and RHIC \cite{Rafelski2015}.
Also, Bodmer and Witten \cite{Bodmer1971,Witten84} pointed out that it is possible the ground state of matter 
at zero pressure and large baryon number is not iron but strange quark matter. Thus from a theoretical standpoint 
one can have pure quark stars. In \cite{Zhou2017xhf,Zhou:2017xkz,Zhou:2018ovt,Zhou2019hyy,Zhou2021s} we have computed 
rotating isolated axisymmetric, triaxial and differentially rotating quark stars, while in \cite{Zhou2021}
irrotational binary quark stars have been evolved in grid based hydrodynamics. 
Here we present an example using the PCOCAL code of a spinning binary quark star, 
where each companion has a spin along the axis of the orbital angular momentum.

As in our previous studies the quark EOS is the MIT model with $p=\frac{1}{3}(\GE-4B)$ or in a parametric form
\beqn
& p =  K \GR^{4/3} - B ,\label{eq:eosP }   \\
& \GE =  3K \GR^{4/3} + B  \label{eq:eose}
\eeqn
where $B=\GE_s/4$ is the bag constant which is set to $52.5\ \rm MeV\ fm^{-3}$, and $\GE_s$ the surface energy density. 
The parameter $\GR$ could be regarded as the rest-mass density  and $K$ is constant. As explained in \cite{Zhou2021s},
changing $K$ does not affect the EOS, thus we choose this constant such that at the surface $\GR_s=\GE_s$ i.e.
$h_s=1$. This is accomplished for 
\beq
K = \left( \frac{c^8}{256 B}\right)^{\frac{1}{3}} = 3.119\times 10^{15}\ \frac{cm^3}{s^2 g^{1/3}}.
\eeq
Assuming $\GR_s = 1.4 \GR_{\rm nuc}$ this EOS predicts a maximum spherical mass of a quark star of
$\maxtov=2.1 M_\odot$ with $R_{\rm max}^{\rm sph}=11.5\ \rm km$.

\begin{figure*}[h]
\centering
\includegraphics[width=0.99\textwidth]{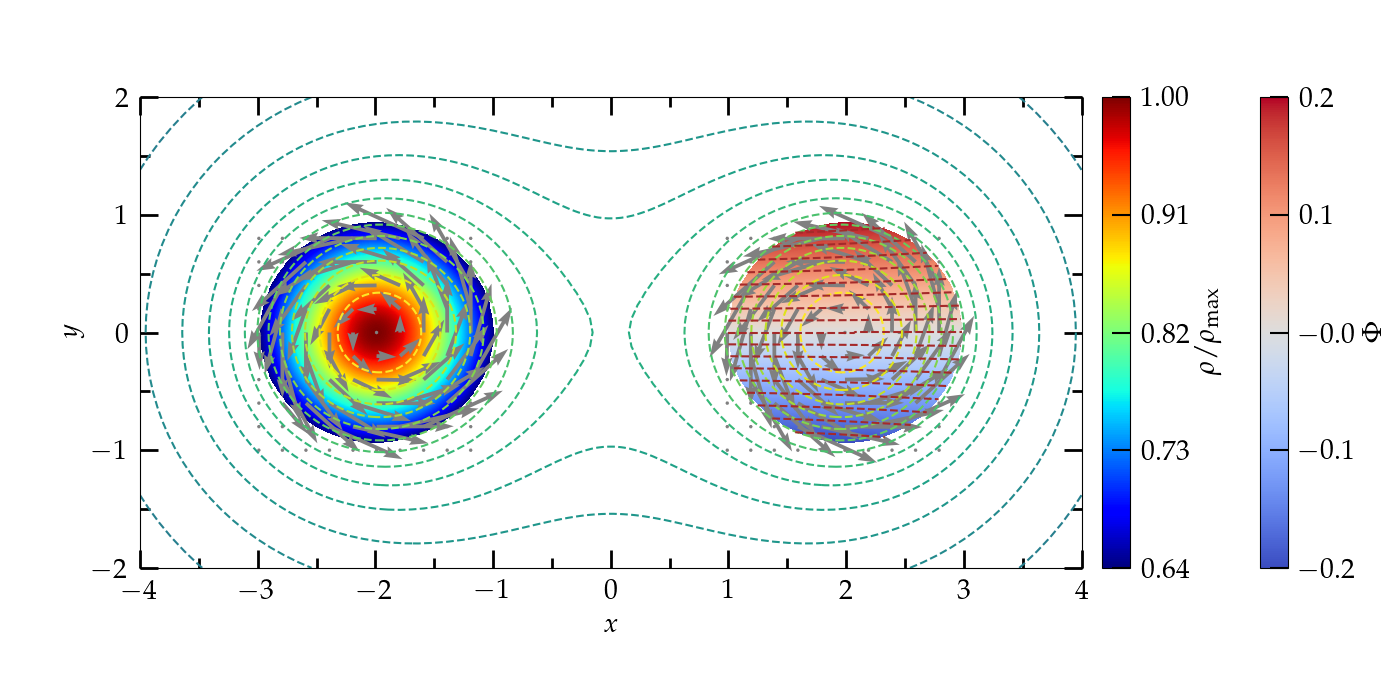}
\caption{A spinning binary quark star. The rest-mass density $\GR$ is shown for the star on the left 
while the velocity potential $\Phi$ is plotted for the star on the right. Contour plots of the conformal 
factor $\GC$  (green dashed lines) and the velocity potential $\Phi$ (brown dashed lines in the right quark star),
as well as the fluid velocity with respect to the corotating observer, are shown.
}
\label{fig:sbqs}
\end{figure*}

Using the formalism described in \ref{sec:bns} with a spin vector
\beq
s^i = 0.13 (-y_s,x_s,0)  \label{eq:spin}
\eeq
($x_s,y_s$ are coordinates with respect to the quark star center) in Eq. (\ref{eq:uih_dec}), and solving Eqs.
(\ref{eq:psi}), (\ref{eq:shift}), (\ref{eq:lapse}) and (\ref{eq:cbm4}), (\ref{eq:uih_dec}), (\ref{eq:irspei}),
we compute here a spinning binary quark star at $37.3\ \rm km$ 
separation and Arnowitt-Deser-Misner mass $M=2.84$. The quasilocal spin \cite{Tsokaros2018} of each
companion is $J_{\rm ql}=0.858$ along the axis of orbital motion. This spin angular momentum corresponds to 
an approximate dimensionless spin of $J_{\rm ql}/(M/2)^2=0.43$.
In Fig. \ref{fig:sbqs} we plot the rest-mass density (left quark star), the velocity potential (right quark star),
contour plots of the conformal factor $\GC$ (green dashed lines) and velocity potential (brown dashed lines), as well
as the fluid velocity with respect to the corotating frame. The direction of the velocity arrows show that the spin
angular momentum is along the positive z axis (contrary to the irrotational case) in accordance with Eq. (\ref{eq:spin}). 
From the density colorbar we see that the central
density $\GR_c$ is approximately $1.55$ times the surface density $\GR_s$. The nonzero surface density is characteristic 
of the self-bound quark stars contrary to their neutron star analogues. For nonspinning (irrotational) binaries of the 
same surface density (and rest mass) the central density is higher and $\GR_c/\GR_s=1.63$. This behavior is similar
to binary neutron stars as well as to isolated rotating neutron/quark stars.
Note that in order to find the quasiequilibrium solution a root finding method over the
central quark star densities is needed, and therefore a number of cycles ($\sim 10$) of Algorithm 2 is performed.
Pointwise comparison between PCOCAL and COCAL solutions show differences $\lesssim 10^{-12}$
for all variables.

\section{Conclusions}
In this work we presented a new efficient parallelized code, PCOCAL, for general relativistic compact object 
initial data. PCOCAL is based on the serial COCAL code (which has been tested for a wide range of atrophysical
scenarios) and employs extensive OpenMP practices. To optimize PCOCAL, a thorough profiling of the serial COCAL code
is done for the first time. 
Both the single and the binary neutron star modules have been examined in detail and timing
of each subroutine was performed. The parallelized module that solves the full set of Einstein equations for a single
rotating neutron star is found to be 12-14 times faster than the corresponding serial code, 
with the parallel part being 18-20 times faster than the serial one. The parallelized module that 
solves for a binary neutron star system in quasiequilibrim is 5-6 times faster than the corresponding serial code,
with the parallel part being 12-16 times faster than the serial one.
As a novel example, we computed initial data for a spinning binary quark star system that could be 
a binary neutron star mimicker.
The achieved speedup will help us explore a larger set of problems that require intensive calculations as for example
in the calculation of magnetized neutron stars. 

As a step forward we plan to revisit the data communication and memory hadling of the binary module in order
to achieve better speedups for the highest resolutions, as in the single rotating star module. In addition 
we will perform domain decomposion in order to implement a combined MPI-OpenMP scheme.

\section{Acknowledgements}
  A.T. thanks E. Zhou for useful discussions. This work was supported in part by National Science Foundation 
  Grants No. PHY-2308242 and No. OAC-2310548 to the University of Illinois at Urbana-Champaign. 
  A.T. acknowledges support from the National Center for Supercomputing Applications (NCSA) at the University of
  Illinois at Urbana-Champaign through the NCSA Fellows program.
  K.U. is supported by JSPS Grant-in-Aid for Scientific Research (C) 22K03636 to the 
  University of the Ryukyus.

\section*{Conflict of interest}
 The authors declare that they have no conflict of interest.

\section*{Declarations}
The authors declare that there are no data associated with this manuscript.

\appendix

\section{Source terms}
\label{sec:st}

The source terms $\mathcal{S}_{\rm H}$, $\mathcal{S}_i$, $\mathcal{S}_{\rm tr}$, and ${\cal S}_{ij}$
in Eqs. (\ref{eq:psi})-(\ref{eq:hab}) are 
\begin{eqnarray}
\mathcal{S}_{\rm H} & = & \color{magenta} \frac{1}{8}\GC\tRs \color{black} 
                        - \color{magenta} h^{ij}\fD_i\fD_j\GC \color{black} 
                        + \color{magenta} \tGG^{ij}\C{k}{i}{j}\fD_k\GC \color{black} \nonumber  \\ 
      & - & \frac{\GC^5}{8}\left(\tA_{ij}\tA^{ij}-\frac{2}{3}K^2\right) - 2\pi \GR_{\rm H} \GC^5 ,  \label{eq:HC} \\[5pt]
\mathcal{S}_i & = & - \color{magenta} \tRs_{ij}\tGB^j \color{black} 
                    - \color{magenta} h^{ab}\fD_a\fD_b\tGB_i   \nonumber  \\
              & + & \color{magenta} \tGG^{ab} [ \fD_a(\C{m}{b}{i}\tGB_m) \color{black} 
                    + \color{magenta} \C{m}{a}{b}\TD{D}{m}\tGB_i \color{black}
                    + \color{magenta} \C{m}{a}{i}\TD{D}{b}\tGB_m \color{black}  ]    \nonumber \\
        & - & \frac{1}{3}\fD_i\fD_j \tGB^j 
                    -\frac{2\GA^2}{\GC^6}\TDU{A}{i}{j}\fD_j\left(\frac{\GC^6}{\GA}\right)  \nonumber \\ 
             & + &   \frac{4\GA}{3}\fD_i K + \tGG^{jm}\TD{D}{j} \tilde{u}_{im} + 16\pi\GA j_i \label{eq:MC} ,   \\[5pt]
\mathcal{S}_{\rm tr} & = &  \color{magenta}  \frac{\GA\GC}{8}\tRs \color{black} 
                           - \color{magenta} h^{ij}\fD_i\fD_j(\GA\GC) \color{black}  
                          + \color{magenta}  \tGG^{ij}\C{k}{i}{j}\fD_k(\GA\GC)   \nonumber   \\
     & + & \GA\GC^5\left(\frac{7}{8}\tA_{ij}\tA^{ij}+\frac{5}{12}K^2\right)    \nonumber  \\
     & - & \GC^5\mathcal{L}_{\GA \Bn}K + 2\pi\GA\GC^5(\GR_{\rm H}+2S) ,  \label{eq:trdKijdt} \\[5pt]
\color{magenta} 
{\cal S}_{ij} & \color{magenta} = & \color{magenta}  - \frac{1}{3}\tGG_{ij}\fD_k h_{ab}\fDu{k}h^{ab} 
                     + \frac{2}{3}\tGG_{ij}\fDu{k}\C{a}{a}{k} \nonumber  \\ 
              & \color{magenta} + & \color{magenta}  2\left[   \tRs_{ij}^{\li} 
                           +  \tRs_{ij}^{\nl} + \GC^4\left(\frac{1}{3}K\tA_{ij}-2\tA_{ik}\TUD{A}{k}{j}\right)\right.   \nonumber \\
             & \color{magenta} + &  \color{magenta}  \frac{1}{\GA\GC^2}\left(-\fD_i\fD_j(\GA\GC^2) +  
                     \C{k}{i}{j}\fD_k(\GA\GC^2)    \right.   \nonumber  \\
             & \color{magenta} + &   \color{magenta}   \left. 4\fD_i(\GA\GC)\fD_j\GC + 4\fD_i\GC\fD_j(\GA\GC)  \right)  \nonumber   \\ 
             & \color{magenta} - & \color{magenta}   \left.\frac{1}{\GA} \mathcal{L}_{\GA\Bn}(\GC^4\tA_{ij}) - 
                                   8\pi S_{ij}  \right]^{\tf} ,  \label{eq:tfdKijdt}
\end{eqnarray}
where with magenta color we signify terms that are due to the nonconformal flat contribution $h^{ij}$. 
Such terms are zero in the IWM formulation used in the binary neutron star calculations of Sec. \ref{sec:bns}.
We also define the conformally rescaled shift as  
 $\TU{\GB}{i}=\GB^i$, therefore $\TD{\GB}{i}=\TDD{\GG}{i}{j}\TU{\GB}{j}=\psi^{-4} \GB_i$.
The covariant derivatives associated with $\GG_{ij},\tGG_{ij}$, and $f_{ij}$ are respectively $D, \tilde{D}$, and $\fD$. 
It is 
\begin{eqnarray}
D_i\GB^k  & = & \tD_i\GB^k + \tilde{C}^k_{\ ij}\GB^j  \\
\tD_i \GB^k & = & \fD_i\GB^k + C^k_{\ ij}\GB^j
\end{eqnarray}
where
\begin{eqnarray}
\tilde{C}^k_{\ ij} & = & \frac{2}{\GC}(\tGG^k_{\ i}\tD_j\GC + \tGG^k_{\ j}\tD_i\GC - \tGG_{ij}\tGG^{km}\tD_m\GC)  \\
C^k_{\ ij} & = & \frac{1}{2}\tGG^{km}(\fD_i h_{mj}+\fD_j h_{mi}-\fD_m h_{ij})
\end{eqnarray}
and 
\begin{equation}
C^k_{\ kj}=\frac{1}{2\tGG}\fD_j\tGG = 0 ,\qquad \tilde{C}^k_{\ kj}=\frac{1}{2\GG}\tD_j\GG
\end{equation}
since $\tGG=f$. 
In the case where Cartesian coordinates are used for the flat metric, i.e. $f_{ij}=\GD_{ij}$, then
$\fD$ is the usual partial derivative $\pd$, and $\flap$ the Laplacian in Cartesian coordinates.
The superscript $^{\tf}$ means the trace-free part.
The conformal tracefree extrinsic curvature is written as
\beq
\tA_{ij} = \frac{1}{2\GA}[ \TBD{i}{j} - \tilde{u}_{ij} ],\qquad  \tilde{u}_{ij} := [\pd_t \tGG_{ij}]^{\tf},
\label{eq:taij}
\eeq
where
$\tilde{\mathbb L}$ is the conformal Killing 
operator:
$\TBD{i}{j}=\TD{D}{i}\TD{\GB}{j}+\TD{D}{j}\TD{\GB}{i}-\frac{2}{3}\TDD{\GG}{i}{j}\TD{D}{k}\TU{\GB}{k}$.

The matter sources that appear on the right-hand side of Eqs. 
(\ref{eq:HC})-(\ref{eq:tfdKijdt}) are  
\begin{eqnarray}
\GR_{\rm H} & := & T_{\mu\nu}n^\mu n^\nu , \\
j_i & := & -T_{\mu\nu}\GG^\mu_{\ \,i} n^\nu ,  \\
S & := & T_{\mu\nu}\GG^{\mu\nu},   \\
S_{ij} & := & T_{\mu\nu}\GG^\mu_{\ \,i}\GG^\nu_{\ \,j}.
\end{eqnarray}

Another important term in our system is the one that involves $\tRs_{ij}$, the 3-d
Ricci tensor associated with the conformal geometry $\tGG_{ij}$. One can show \cite{Shibata04} that
\beq
\tRs_{ij} = -\frac{1}{2} \flap h_{ij} + \tRs_{ij}^{\li} + \tRs_{ij}^{\nl} , \label{eq:tRij}
\eeq
where 
\begin{eqnarray}
\tRs_{ij}^{\li} & := &  \frac{1}{2}(f_{ik}\fD_j \TU{\Gamma}{k} + f_{jk}\fD_i \TU{\Gamma}{k}),  \label{eq:tRks}  \\
\tRs_{ij}^{\nl} & := & -\frac{1}{2}(h^{ab}\fD_a\fD_b h_{ij} + \fD_i h^{ab}\fD_b h_{aj} + \fD_j h^{ab} \fD_a h_{ib}) \nonumber \\ 
                & + &  \frac{1}{2}[\fD_i(h_{kj} \TU{\Gamma}{k}) + \fD_j(h_{ik} \TU{\Gamma}{k})]  \nonumber  \\
                & - & \fD_i C^k_{\ kj} + C^k_{\ km}C^m_{\ ij} - \TU{\Gamma}{k} C_{kij} - C^k_{\ im}C^m_{\ kj} ,  \label{eq:rijnl}
\end{eqnarray}
and 
\beq
\TU{\Gamma}{i} := - \fD_j \tGG^{ij} = - \fD_j h^{ij}
\label{eq:fi}
\eeq
Notice that the terms $\tRs_{ij}^{\li}$,  $\tRs_{ij}^{\nl}$ enter into Eq. (\ref{eq:tfdKijdt}), which we
discuss below. The nonlinear term $\tRs_{ij}^{\nl}$ is second order in $h_{ij}$ and therfore smaller than
the first order terms $\tRs_{ij}^{\li}$ and $\flap h_{ij}$ in Eq.
(\ref{eq:tRij}). The term $\tRs_{ij}^{\li}$
involves the gauge functions $\TU{\Gamma}{i}$ which are the conformal connection functions in the BSSN formulation
\cite{Nakamura87,Shibata95,Baumgarte99}. For initial data, in order for the 
whole system (\ref{eq:HC}-\ref{eq:tfdKijdt}) to converge, these functions must be fixed \cite{Shibata04,Tsokaros2018a}
\beq
\TU{\Gamma}{i} = -H^i \ 
\label{eq:gaucon}
\eeq
where $H^i$ is known.
Eqs. (\ref{eq:gaucon}), are related to our freedom in choosing spatial coordinates. 
In order to impose Eqs. (\ref{eq:gaucon}) to the solutions of the system Eq. (\ref{eq:HC})-(\ref{eq:tfdKijdt}),
and thereby having a self-consistent iteration scheme, an adjustment is necessary for the $h_{ij}$.
Following \cite{Uryu:2009ye}, (or \cite{Uryu2016a} Eq. (29-32)) gauge vector potentials $\xi^a$ are introduced through the 
transformation 
\beq
\GD \GG^{ij} \rightarrow \GD \GG^{ij} \,-\, \fDu{i} \xi^j \,-\, \fDu{j} \xi^i, 
\label{eq:gauge_transf}
\eeq
which are then used to adjust $h^{ij}$ as 
\beq
h^{ij}{}' =  h^{ij} \,-\, \fDu{i}\xi^j \,-\, \fDu{j}\xi^i \,+\, \frac{2}{3}f^{ij} \fD_k\xi^k \, .
\label{eq:gauge_hab}
\eeq
Here $h^{ij}{}'$ are chosen to satisfy the condition $\fD_i h^{ij}{}'=H^j$ given by (\ref{eq:fi}).  
The gauge vector potentials $\xi^i$ are then solved from the elliptic equations, 
\beq
 \flap \xi^i \,=\, \fD_j h^{ij} - \frac{1}{3} \fDu{i} \fD_j\xi^j - H^i ,
\label{eq:gauge}
\eeq
and subsequently the $h_{ij}$ are replaced by Eq.~(\ref{eq:gauge_hab}).

\bibliographystyle{spphys}       

\bibliography{atreferences}

\end{document}

%% file: defs.tex
\newcommand{\beq}{\begin{equation}} 
\newcommand{\eeq}{\end{equation}} 
\newcommand{\beqn}{\begin{eqnarray}} 
\newcommand{\eeqn}{\end{eqnarray}} 

\newcommand{\GA}{\alpha}
\newcommand{\GB}{\beta}
\newcommand{\GG}{\gamma}
\newcommand{\GD}{\delta}
\newcommand{\GE}{\epsilon}

\newcommand{\GL}{\lambda}
\newcommand{\GR}{\rho}

\newcommand{\GC}{\psi}

\newcommand{\GO}{\omega}

\newcommand{\GP}{\phi}

\newcommand{\GU}{\theta}
\newcommand{\pd}{\partial}
\newcommand{\na}{\nabla}

\newcommand{\Bn}{\mathbf{n}}                                                                                                                         
\newcommand{\Bt}{\mathbf{t}}
\newcommand{\Bk}{\mathbf{k}}
\newcommand{\Bbe}{\boldsymbol{\GB}}

\newcommand{\Bom}{\boldsymbol{\GO}}
                                                                                                                                                    
\newcommand{\TD}[2]{\tilde{#1}_{#2}}
\newcommand{\TU}[2]{\tilde{#1}^{#2}}
\newcommand{\TUU}[3]{\tilde{#1}^{#2 #3}}
\newcommand{\TDD}[3]{\tilde{#1}_{#2 #3}}
\newcommand{\TDU}[3]{\tilde{#1}_{#2}^{\ #3}}
\newcommand{\TUD}[3]{\tilde{#1}^{#2}_{\ #3}}

\newcommand{\TBD}[2]{( \tilde{\mathbb L} \tilde{\GB} )_{#1 #2}}

\newcommand{\tA}{\tilde{A}}
\newcommand{\tD}{\tilde{D}}

\newcommand{\tGG}{\tilde{\GG}}
\newcommand{\tGB}{\tilde{\GB}}

\newcommand{\tRs}{\tilde{\mathcal{R}}}

\newcommand{\C}[3]{C^{#1}_{\ {#2}{#3}}}

\newcommand{\flap}{\stackon[0.07ex]{$\Delta$}{\scalebox{0.7}{$\circ$}}}
\newcommand{\fDu}[1]{\stackon[-0.3ex]{$D^{#1}$}{\kern-0.5ex\scalebox{0.7}{$\circ$}}}
\newcommand{\fD}{\stackon[0.07ex]{$D$}{\scalebox{0.7}{$\circ$}}}

\newcommand{\alone}{{\raise0.5ex\hbox{${}^{\ 1}$}}\!\!\!\!\alpha}

\newcommand{\nalam}{\mathrel{\raise0.9ex\hbox{$^\lambda$}\mkern-14mu
\lower0.0ex\hbox{$\nabla$}}}

\newcommand{\zeroD}{{\raise1.0ex\hbox{${}^{\ \circ}$}}\!\!\!\!\!D}
\newcommand{\zna}{{\raise1.0ex\hbox{${}^{\ \circ}$}}\!\!\!\!\!\nabla}
\newcommand{\zS}{{\raise1.0ex\hbox{${}^{\ \circ}$}}\!\!\!\!\!S}

\newcommand{\nl}{\rm \scriptscriptstyle NL}
\newcommand{\li}{\rm \scriptscriptstyle LI}
\newcommand{\tf}{\rm \scriptscriptstyle TF}

\newcommand{\Nrf}{{N_r^{\rm f}}}
\newcommand{\Nrm}{{N_r^{\rm m}}}

\newcommand{\cocal}{\textsc{cocal}}

\newcommand{\maxtov}{M_{\rm max}^{\rm sph}}

%% file: pcocal.bbl
\begin{thebibliography}{10}
\providecommand{\url}[1]{{#1}}
\providecommand{\urlprefix}{URL }
\expandafter\ifx\csname urlstyle\endcsname\relax
  \providecommand{\doi}[1]{DOI \discretionary{}{}{}#1}\else
  \providecommand{\doi}{DOI \discretionary{}{}{}\begingroup
  \urlstyle{rm}\Url}\fi

\bibitem{Abbot2016-GW-detection-prl}
B.P. {Abbott}, R.~{Abbott}, T.D. {Abbott}, M.R. {Abernathy}, F.~{Acernese},
  K.~{Ackley}, C.~{Adams}, T.~{Adams}, P.~{Addesso}, R.X. {Adhikari}, et~al.,
  Phys. Rev. Lett. \textbf{116}(6), 061102 (2016).
\newblock \doi{10.1103/PhysRevLett.116.061102}

\bibitem{GW170817prl}
B.P. Abbott, et~al., Phys. Rev. Lett. \textbf{119}(16), 161101 (2017).
\newblock \doi{10.1103/PhysRevLett.119.161101}

\bibitem{Kolzova2017circ}
A.~{Kozlova}, S.~{Golenetskii}, R.~{Aptekar}, D.~{Frederiks}, D.~{Svinkin},
  T.~{Cline}, K.~{Hurley}, V.~{Connaughton}, M.S. {Briggs}, C.~{Meegan},
  V.~{Pelassa}, A.~{Goldstein}, A.~{von Kienlin}, X.~{Zhang}, A.~{Rau},
  V.~{Savchenko}, E.~{Bozzo}, C.~{Ferrigno}, S.~{Barthelmy}, J.~{Cummings},
  H.~{Krimm}, D.~{Palmer}, GRB Coordinates Network, Circular Service,
  No.~21517, \#1 (2017) \textbf{1517} (2017)

\bibitem{Abbott2017b}
B.P. Abbott, et~al., Astrophys. J. Lett. \textbf{848}(2), L12 (2017).
\newblock \doi{10.3847/2041-8213/aa91c9}

\bibitem{Abbott2017d_etal}
B.P. Abbott, et~al., Astrophys. J. Lett. \textbf{848}(2), L13 (2017).
\newblock \urlprefix\url{http://stacks.iop.org/2041-8205/848/i=2/a=L13}

\bibitem{Abbott2017c}
B.P. {Abbott}, et~al., Astrophys. J. Lett. \textbf{850}, L39 (2017).
\newblock \doi{10.3847/2041-8213/aa9478}

\bibitem{Huang08}
X.~Huang, C.~Markakis, N.~Sugiyama, K.~Ury{\={u}}, Phys. Rev. D \textbf{D78},
  124023 (2008).
\newblock \doi{10.1103/PhysRevD.78.124023}

\bibitem{Uryu2016a}
K.~Ury{\={u}}, A.~Tsokaros, F.~Galeazzi, H.~Hotta, M.~Sugimura, K.~Taniguchi,
  S.~Yoshida, Phys. Rev. D \textbf{D93}(4), 044056 (2016).
\newblock \doi{10.1103/PhysRevD.93.044056}

\bibitem{Uryu2016b}
K.~{Ury{\=u}}, A.~{Tsokaros}, L.~{Baiotti}, F.~{Galeazzi}, N.~{Sugiyama},
  K.~{Taniguchi}, S.~{Yoshida}, Phys. Rev. D \textbf{94}(10), 101302 (2016).
\newblock \doi{10.1103/PhysRevD.94.101302}

\bibitem{Uryu2017}
K.~Uryu, A.~Tsokaros, L.~Baiotti, F.~Galeazzi, K.~Taniguchi, S.~Yoshida, Phys.
  Rev. D \textbf{96}(10), 103011 (2017).
\newblock \doi{10.1103/PhysRevD.96.103011}

\bibitem{Zhou2017xhf}
E.~Zhou, A.~Tsokaros, L.~Rezzolla, R.~Xu, K.~Uryū, Phys. Rev. \textbf{D97}(2),
  023013 (2018).
\newblock \doi{10.1103/PhysRevD.97.023013}

\bibitem{Zhou2019hyy}
E.~Zhou, A.~Tsokaros, K.~Uryu, R.~Xu, M.~Shibata, Phys. Rev. \textbf{D100}(4),
  043015 (2019).
\newblock \doi{10.1103/PhysRevD.100.043015}

\bibitem{Uryu2012}
K.~{Ury{\={u}}}, A.~{Tsokaros}, Phys. Rev. D \textbf{85}(6), 064014 (2012).
\newblock \doi{10.1103/PhysRevD.85.064014}

\bibitem{Tsokaros2012}
A.~Tsokaros, K.~Ury{\={u}}, Journal of Engineering Mathematics \textbf{82}(1),
  133 (2012).
\newblock \doi{10.1007/s10665-012-9585-6}.
\newblock \urlprefix\url{http://dx.doi.org/10.1007/s10665-012-9585-6}

\bibitem{Uryu:2012b}
K.~Ury{\={u}}, A.~Tsokaros, P.~Grandclement, Phys. Rev. D \textbf{D86}, 104001
  (2012).
\newblock \doi{10.1103/PhysRevD.86.104001}

\bibitem{Tsokaros2015}
A.~{Tsokaros}, K.~{Ury{\={u}}}, L.~{Rezzolla}, Phys. Rev. D \textbf{91}(10),
  104030 (2015).
\newblock \doi{10.1103/PhysRevD.91.104030}

\bibitem{Tsokaros2018}
A.~Tsokaros, K.~Uryu, M.~Ruiz, S.L. Shapiro, Phys. Rev. \textbf{D98}(12),
  124019 (2018).
\newblock \doi{10.1103/PhysRevD.98.124019}

\bibitem{Uryu2014}
K.~Uryu, E.~Gourgoulhon, C.~Markakis, K.~Fujisawa, A.~Tsokaros, Y.~Eriguchi,
  Phys. Rev. \textbf{D90}(10), 101501 (2014).
\newblock \doi{10.1103/PhysRevD.90.101501}

\bibitem{Uryu2019}
K.~Uryu, S.~Yoshida, E.~Gourgoulhon, C.~Markakis, K.~Fujisawa, A.~Tsokaros,
  K.~Taniguchi, Y.~Eriguchi, Phys. Rev. \textbf{D100}(12), 123019 (2019).
\newblock \doi{10.1103/PhysRevD.100.123019}

\bibitem{Uryu:2023lgp}
K.~Uryu, S.~Yoshida, E.~Gourgoulhon, C.~Markakis, K.~Fujisawa, A.~Tsokaros,
  K.~Taniguchi, M.~Zamani, Phys. Rev. D \textbf{107}(10), 103016 (2023).
\newblock \doi{10.1103/PhysRevD.107.103016}

\bibitem{Tsokaros2018a}
A.~Tsokaros, K.~Uryu, S.L. Shapiro, Phys. Rev. \textbf{D99}(4), 041501 (2019).
\newblock \doi{10.1103/PhysRevD.99.041501}

\bibitem{Uryu00}
K.~{Ury{\=u}}, Y.~{Eriguchi}, Phys. Rev. D \textbf{61}(12), 124023 (2000).
\newblock \doi{10.1103/PhysRevD.61.124023}

\bibitem{Tsokaros2007}
A.A. Tsokaros, K.~Ury{\={u}}, Phys. Rev. D \textbf{75}, 044026 (2007).
\newblock \doi{10.1103/PhysRevD.75.044026}.
\newblock \urlprefix\url{http://link.aps.org/doi/10.1103/PhysRevD.75.044026}

\bibitem{Komatsu89}
H.~{Komatsu}, Y.~{Eriguchi}, I.~{Hachisu}, Mon. Not. R. Astron. Soc.
  \textbf{237}, 355 (1989).
\newblock \doi{10.1093/mnras/237.2.355}

\bibitem{Komatsu89b}
H.~{Komatsu}, Y.~{Eriguchi}, I.~{Hachisu}, Mon. Not. R. Astron. Soc.
  \textbf{239}, 153 (1989)

\bibitem{Ansorg:2004ds}
M.~Ansorg, B.~Br{\"u}gmann, W.~Tichy, Phys. Rev. D \textbf{70}, 064011 (2004)

\bibitem{lorene}
\urlprefix\url{http://www.lorene.obspm.fr}.
\newblock Langage Objet pour la RElativit\'{e} Num\'{e}rique,
  \url{www.lorene.obspm.fr}

\bibitem{kadath}
\urlprefix\url{https://kadath.obspm.fr}

\bibitem{Grandclement09}
P.~Grandclement, J. Comput. Phys. \textbf{229}, 3334 (2010).
\newblock \doi{10.1016/j.jcp.2010.01.005}

\bibitem{Papenfort2021}
L.J. Papenfort, S.D. Tootle, P.~Grandcl\'ement, E.R. Most, L.~Rezzolla, Phys.
  Rev. D \textbf{104}(2), 024057 (2021).
\newblock \doi{10.1103/PhysRevD.104.024057}

\bibitem{Tichy:2009}
W.~Tichy, Class. Quant. Grav. \textbf{26}, 175018 (2009).
\newblock \doi{10.1088/0264-9381/26/17/175018}

\bibitem{Pfeiffer:2002wt}
H.P. Pfeiffer, L.E. Kidder, M.A. Scheel, S.A. Teukolsky, Comput. Phys. Commun.
  \textbf{152}, 253 (2003)

\bibitem{Rashti2021}
A.~Rashti, F.M. Fabbri, B.~Br\"ugmann, S.V. Chaurasia, T.~Dietrich, M.~Ujevic,
  W.~Tichy, Phys. Rev. D \textbf{105}(10), 104027 (2022).
\newblock \doi{10.1103/PhysRevD.105.104027}

\bibitem{Assumpcao2021}
T.~Assumpcao, L.R. Werneck, T.P. Jacques, Z.B. Etienne, Phys. Rev. D
  \textbf{105}(10), 104037 (2022).
\newblock \doi{10.1103/PhysRevD.105.104037}

\bibitem{Tsokaros2022r}
A.~Tsokaros, K.~Ury\={u}, Gen. Rel. Grav. \textbf{54}(6), 52 (2022).
\newblock \doi{10.1007/s10714-022-02928-1}

\bibitem{Isenberg08}
J.A. {Isenberg}, International Journal of Modern Physics D \textbf{17}, 265
  (2008).
\newblock \doi{10.1142/S0218271808011997}

\bibitem{Wilson89}
J.R. {Wilson}, G.J. {Mathews}, \emph{{Relativistic hydrodynamics.}} (Cambridge
  University Press, 1989), pp. 306--314

\bibitem{Wilson95}
J.R. Wilson, G.J. Mathews, Phys. Rev. Lett. \textbf{75}, 4161 (1995)

\bibitem{Wilson96}
J.R. {Wilson}, G.J. {Mathews}, P.~{Marronetti}, Phys. Rev. D \textbf{54}, 1317
  (1996).
\newblock \doi{10.1103/PhysRevD.54.1317}

\bibitem{Shibata04a}
M.~Shibata, K.~Uryu, J.L. Friedman, Phys. Rev. \textbf{D70}, 044044 (2004).
\newblock \doi{10.1103/PhysRevD.70.044044, 10.1103/PhysRevD.70.129901}.
\newblock [Erratum: Phys. Rev.D70,129901(2004)]

\bibitem{Bonazzola:2003dm}
S.~Bonazzola, E.~Gourgoulhon, P.~Grandclement, J.~Novak, Phys. Rev. D
  \textbf{70}, 104007 (2004)

\bibitem{Uryu2006}
K.~{Ury{\=u}}, F.~{Limousin}, J.L. {Friedman}, E.~{Gourgoulhon}, M.~{Shibata},
  Phys. Rev. Lett. \textbf{97}(17), 171101 (2006).
\newblock \doi{10.1103/PhysRevLett.97.171101}

\bibitem{Uryu:2009ye}
K.~Ury{\=u}, F.~Limousin, J.L. Friedman, E.~Gourgoulhon, M.~Shibata, Phys. Rev.
  D \textbf{80}, 124004 (2009).
\newblock \doi{10.1103/PhysRevD.80.124004}

\bibitem{Bonazzola97}
S.~{Bonazzola}, E.~{Gourgoulhon}, J.A. {Marck}, Phys. Rev. D \textbf{56}, 7740
  (1997).
\newblock \doi{10.1103/PhysRevD.56.7740}

\bibitem{Asada1998}
H.~{Asada}, Phys. Rev. D \textbf{57}, 7292 (1998).
\newblock \doi{10.1103/PhysRevD.57.7292}

\bibitem{Shibata98}
M.~{Shibata}, Phys. Rev. D \textbf{58}(2), 024012 (1998).
\newblock \doi{10.1103/PhysRevD.58.024012}

\bibitem{Teukolsky98}
S.A. {Teukolsky}, Astrophys. J. \textbf{504}, 442 (1998).
\newblock \doi{10.1086/306082}

\bibitem{Tichy11}
W.~{Tichy}, Phys. Rev. D \textbf{84}(2), 024041 (2011).
\newblock \doi{10.1103/PhysRevD.84.024041}

\bibitem{Rafelski2015}
J.~Rafelski, Eur. Phys. J. A \textbf{51}(9), 114 (2015).
\newblock \doi{10.1140/epja/i2015-15114-0}

\bibitem{Bodmer1971}
A.R. {Bodmer}, Phys. Rev. D \textbf{4}, 1601 (1971).
\newblock \doi{10.1103/PhysRevD.4.1601}

\bibitem{Witten84}
E.~Witten, Phys. Rev. D \textbf{30}, 272 (1984).
\newblock \doi{10.1103/physrevd.30.272}

\bibitem{Zhou:2017xkz}
E.~Zhou, A.~Tsokaros, L.~Rezzolla, R.~Xu, Astron. Nachr. \textbf{338}(9-10),
  1044 (2017).
\newblock \doi{10.1002/asna.201713432}

\bibitem{Zhou:2018ovt}
E.~Zhou, A.~Tsokaros, L.~Rezzolla, R.~Xu, K.~Ury\={u}, Universe \textbf{4}(3),
  48 (2018).
\newblock \doi{10.3390/universe4030048}

\bibitem{Zhou2021s}
E.~Zhou, K.~Kiuchi, M.~Shibata, A.~Tsokaros, K.~Uryu, Phys. Rev. D
  \textbf{103}, 123011 (2021).
\newblock \doi{10.1103/PhysRevD.103.123011}

\bibitem{Zhou2021}
E.~Zhou, K.~Kiuchi, M.~Shibata, A.~Tsokaros, K.~Uryu, Phys. Rev. D
  \textbf{106}(10), 103030 (2022).
\newblock \doi{10.1103/PhysRevD.106.103030}

\bibitem{Shibata04}
M.~{Shibata}, Y.~{Sekiguchi}, Phys. Rev. D \textbf{69}(8), 084024 (2004)

\bibitem{Nakamura87}
T.~{Nakamura}, K.~{Oohara}, Y.~{Kojima}, Progress of Theoretical Physics
  Supplement \textbf{90}, 1 (1987).
\newblock \doi{10.1143/PTPS.90.1}

\bibitem{Shibata95}
M.~{Shibata}, T.~{Nakamura}, Phys. Rev. D \textbf{52}, 5428 (1995).
\newblock \doi{10.1103/PhysRevD.52.5428}

\bibitem{Baumgarte99}
T.W. {Baumgarte}, S.L. {Shapiro}, Phys. Rev. D \textbf{59}(2), 024007 (1999).
\newblock \doi{10.1103/PhysRevD.59.024007}

\end{thebibliography}
